\pdfoutput=1
\documentclass[aip,pof,preprint,graphicx,amsmath,amssymb,nofootinbib]{revtex4-1}

\usepackage{pgfplots}
\pgfplotsset{compat=newest}
\usetikzlibrary{calc}
\usetikzlibrary{plotmarks}
\usetikzlibrary{patterns}
\usetikzlibrary{arrows}
\usepgfplotslibrary{colormaps}

\begin{document}

\title{Hybrid POD-FFT analysis of nonlinear evolving coherent structures of DNS wavepacket in laminar-turbulent transition} 

\author{Kean Lee \surname{Kang}}
\affiliation{NUS Graduate School for Integrative Sciences and Engineering, National University of Singapore, Singapore 117456}
\author{K. S. \surname{Yeo}}
\email[]{mpeyeoks@nus.edu.sg\\
This article may be downloaded for personal use only. Any other use requires prior permission of the author and AIP Publishing. The following article appeared in K. L. Kang and K. S. Yeo, Phys. Fluids \textbf{29}, 084105 (2017) and may be found at http://dx.doi.org/10.1063/1.4999348.}
\affiliation{Department of Mechanical Engineering, National University of Singapore, Singapore 117576}

\date{\today}

\begin{abstract}
This paper concerns the study of direct numerical simulation (DNS) data of a wavepacket in laminar turbulent transition in a Blasius boundary layer. The decomposition of this wavepacket into a set of ``modes'' (a basis that spans an approximate solution space) can be achieved in a wide variety of ways. Two well-known tools are the fast Fourier transform (FFT) and the proper orthogonal decomposition (POD). To synergize the strengths of both methods, a hybrid POD-FFT is pioneered, using the FFT as a tool for interpreting the POD modes. The POD-FFT automatically identifies well-known fundamental, subharmonic and Klebanoff modes in the flow, even though it is blind to the underlying physics. Moreover, the POD-FFT further separates the subharmonic content of the wavepacket into three fairly distinct parts: a positively detuned mode resembling a Lambda-vortex, a Craik-type tuned mode and a Herbert-type positive-negative detuned mode pair, in decreasing order of energy. This distinction is less widely recognized, but it provides a possible explanation for the slightly positively detuned subharmonic mode often observed in previous experiments and simulations.
\end{abstract}


\maketitle 

\section{\label{intro} Introduction}
In this paper, we will attempt to bring a modern tool - the proper orthogonal decomposition (POD), to revisit a classical model of laminar turbulent transition - the wavepacket in a Blasius boundary layer. While the output of the POD is in itself rather abstract, we will use the fast Fourier transform as a bridge to the classical theories. To that end, we will begin with a brief introductory recap of some key theories applicable to wavepackets in transition, before explaining our methods for direct numerical simulation (DNS) of the wavepacket and for fast Fourier transforms (FFT) and POD of the DNS data. Thereafter, we will discuss the results and make a case for how the POD both reaffirms the classical theories of resonance and brings something new to the discussion. 

\subsection{\label{subsec:linear-analysis} Linear analysis}
The equation describing the motion of fluids is the Navier-Stokes equation. For incompressible fluids, it may be written as
\begin{gather}
\frac{\partial\mathbf{V}}{\partial t}+(\mathbf{V}\centerdot\nabla)\mathbf{V} =-\nabla\Pi+\frac{1}{\text{Re}}\nabla^2 \mathbf{V},
\label{eq:navier-stokes-momentum}\\
\nabla\centerdot\mathbf{V}=0,
\label{eq:navier-stokes-continuity}
\end{gather}
where $\mathbf{V}$ is the velocity vector field, $\Pi$ is the pressure scalar field, $t$ is time and $\text{Re}$ is the Reynolds number. $\nabla$ is the gradient operator, and $\nabla^2\equiv\nabla\centerdot\nabla$ is the Laplace operator. In general, velocity and pressure are a function of both the spatial position vector $\mathbf{x}$ and time $t$. A common approach is to separate the steady-state (time-independent) basic flow solution $(\mathbf{U}(\mathbf{x}),P(\mathbf{x}))$ from the time-dependent perturbation solution $(\mathbf{u}(\mathbf{x},t),p(\mathbf{x},t))$. For notational simplicity, we do not write the dependence on $\mathbf{x}$ and $t$ explicitly, and so 
\begin{equation}
(\mathbf{V},\Pi)= (\mathbf{U}+\mathbf{u},P+p).
\label{eq:perturbation-split-simplified}
\end{equation}

By substituting (\ref{eq:perturbation-split-simplified}) into the incompressible Navier-Stokes equation (\ref{eq:navier-stokes-momentum}), we can obtain the general flow stability equation that governs the evolution of a velocity perturbation vector $\mathbf{u}$ to a basic flow vector $\mathbf{U}$,
\begin{equation}
\frac{\partial\mathbf{u}}{\partial t}+(\mathbf{U}\centerdot\nabla)\mathbf{u} +(\mathbf{u}\centerdot\nabla)\mathbf{U} +(\mathbf{u}\centerdot\nabla)\mathbf{u} =-\nabla p + \frac{1}{\text{Re}}\nabla^2 \mathbf{u}, 
\label{eq:general-flow-stability}
\end{equation}

For very small perturbations, the nonlinear term $(\mathbf{u}\centerdot\nabla)\mathbf{u}$ can be ignored, giving rise to the linear stability equation, which forms the core of linear stability theory (LST). LST is a popular tool of stability analysis because it is easily amenable to theoretical analysis and provides a sufficient, though not necessary condition for instability. 

If we model the perturbation as a traveling wave, we have
\begin{equation}
(\mathbf{u},p)=(\tilde{\mathbf{u}},\tilde{p}) \exp\left[i\left(\mathbf{k}\centerdot\mathbf{x} -\omega t \right)\right],
\label{eq:traveling-wave-solution}
\end{equation}
where $\mathbf{k}$ is the wavenumber vector, $\omega$ is the angular frequency and $\tilde{\mathbf{u}}$, $\tilde{p}$ are functions describing the amplitude of the traveling wave. By the famous Squire transformation,\cite{Squire1933} any three-dimensional (3D) linear flow stability problem for parallel flow can be converted into an equivalent two-dimensional (2D) case. Squire's transformation thus provided justification for the focus on 2D traveling wave solutions. As a result, to find the minimum critical Reynolds number (the lowest Reynolds number for which the flow becomes unstable), it is sufficient to consider only 2D disturbances. Applying the assumption of 2D traveling wave solutions leads to what is arguably the most well-known equation of flow stability analysis - the Orr-Sommerfeld (OS) equation.\cite{Schmid2001}

The values of phase speed $c$ for which non-trivial solutions of the OS equation exist are known as the eigenvalues of the OS equation. For any given Reynolds number, $\text{Re}$ and wavenumber $\alpha$, there may be many eigenvalues $c$ that satisfy the OS equation. In the inviscid limit of $\text{Re}\rightarrow\infty$, the Orr-Sommerfeld equation approaches the Rayleigh equation, which has a singularity at points in the flow where the perturbation's phase speed is equal to the mean flow velocity, $c=U$. These are known as critical points. Although the singularities will not be present if there is viscosity, it is still found that the flow exhibits special behavior at critical points. In a boundary layer, the points where $c=U$ form critical layers in which strong nonlinear interactions occur. 

There are a wide variety of ways for perturbations to be introduced into the boundary layer, and they are studied in an active field of research now known as boundary layer receptivity.\cite{Morkovin1969} Receptivity is a fascinating process which is far from straightforward, and one of the first experiments on the topic is by Schubauer and Skramstad,\cite{Schubauer1947} who confirmed the existence of largely two-dimensional Tollmien-Schlichting (TS) waves that had been predicted by stability theory. Subsequently, Klebanoff, Tidstrom and Sargent put forward a new perspective that takes into account three-dimensionality in connection with boundary layer instability.\cite{Klebanoff1962} More contemporary work includes studies by Fasel of the interaction between Klebanoff modes and TS waves,\cite{Fasel2002} and by Refs.~\onlinecite{Sengupta2011,Sengupta2012,Bhaumik2014a}, with the latter group reporting an entire transition process from a receptivity stage to a fully turbulent flow.

Once disturbances are successfully introduced into the boundary layer, either from internal sources such as surface roughness and vibrations, or external sources like freestream turbulence,\cite{Saric2002} the perturbations can grow through various linear or nonlinear mechanisms. 

\subsection{\label{subsec:nonlinear-analysis-amplification} Nonlinear analysis: wave amplification} 
If a flow perturbation grows to a stage where the small-amplitude assumption of linear theory is no longer valid, the analysis must be extended to include the nonlinear terms in the general flow stability equation (\ref{eq:general-flow-stability}). Moreover, under some circumstances, several discrete and continuous Orr-Sommerfeld modes\cite{Grosch1978} can interact nonlinearly to trigger transition to turbulence, even though any single mode, if left by itself, is unable to cause transition.\cite{Liu2008}

An important nonlinear mechanism in the development of a Blasius boundary layer perturbation is the Craik triad.\cite{Craik1971} If we use the symbols $\omega$, $\beta$ and $\alpha$ to represent the angular frequency, spanwise and streamwise wavenumber respectively, the Craik triad comprises a 2D fundamental wave mode denoted by subscript $f$, and a symmetric pair of 3D oblique subharmonic waves propagating at equal and opposite angles from the streamwise direction, which we shall denote with the $+$ and $-$ subscripts.
\begin{equation}
\mathbf{v}_f=(\omega,0,\alpha), \quad
\mathbf{v}_+=\left(\frac{\omega}{2},\beta,\frac{\alpha}{2}\right), \quad \mathbf{v}_-=\left(\frac{\omega}{2},-\beta,\frac{\alpha}{2}\right).
\label{eq:craik-triad-components}
\end{equation}
These three modes satisfy the wave resonance condition $\mathbf{v}_f=\mathbf{v}_+ + \mathbf{v}_-$. All three waves have the same phase speed in the streamwise direction, $c=\omega/\alpha$, so the height in the boundary layer at which their phase speed matches the mean flow velocity is the same. As mentioned in the previous section \ref{subsec:linear-analysis}, such a location is known as the critical layer. At this location, an extremely strong nonlinear energy transfer mechanism operates to drive the growth of the subharmonic waves. If the fundamental and subharmonic waves are of the same amplitude, the subharmonics may experience growth an order of magnitude larger than the fundamental.

Further analysis of the subharmonic route to transition was carried out by Herbert.\cite{Herbert1984,Herbert1988} This approach considers the subharmonic as a secondary instability in the boundary layer; it is secondary in the sense that it is a small 3D perturbation riding on a large pre-existing 2D primary disturbance. It differs from Craik's theory because here the 3D perturbation is assumed small relative to the 2D perturbation, whereas in a Craik triad all three modes may be of similar size. Furthermore, all three waves in a Craik triad interact with each other, but in Herbert's secondary instability, the 3D perturbations do not interact with each other, and the 2D wave (part of a periodic mean flow) plays a catalytic role, because its presence influences the growth rates of the 3D modes, but the 2D fundamental itself is not affected by the 3D modes.

While these assumptions theoretically restrict Herbert's theory to lower amplitudes of the subharmonic and hence to earlier stages in the development of a flow disturbance, it is found to be better able to explain two key experimental observations: detuned modes and staggered $\Lambda$-vortices. Firstly, detuned modes become apparent when the solution for the 3D perturbation is converted from the reference frame moving with the 2D perturbation back to the laboratory (fixed) frame. These detuned modes possess frequencies and wavenumbers that do not precisely meet the Craik resonance conditions. Instead, they occur in pairs which are symmetric with respect to the Craik subharmonic mode frequency. If $(\omega,\beta)$ is the frequency and spanwise wavenumber of the fundamental wave, the detuned subharmonic modes occur at $(\frac{1}{2}\omega\pm\Delta\omega,\frac{1}{2}\beta\pm\Delta\beta)$. The phenomena of conjugate detuning in the subharmonic frequency has been observed in the experiments of Kachanov and Levchenko.\cite{Kachanov1984}

Secondly, the spatially periodic part of the solution proposed by Herbert could help explain the staggered vortex configurations observed in smoke visualization experiments by Saric.\cite{Saric1984} These were linked to the subharmonic modes having a component that was invariant to the spatial translations $\mathbf{u}(x,z)=\mathbf{u}(x+2\lambda_x,z+\lambda_z)$, where $x$ and $z$ are the streamwise and spanwise locations respectively, while $\lambda_x$ and $\lambda_z$ are their spatial periods. Incidentally, Herbert's theory also admits a fundamental mode arising from primary resonance between the 2D fundamental and 3D wave system, and which is invariant to the translation $\mathbf{u}(x,z)=\mathbf{u}(x+\lambda_x,z+\lambda_z)$. This regime is characterized by aligned vortices in the streamwise direction. In general, the arrangement of the so-called $\Lambda$-structures in the flow can form a basis for characterization of the transition route in a boundary layer.\cite{Boiko2012} The regime with the subharmonic mode and staggered $\Lambda$-structures is known as the N-type (Novosibirsk) or H-type (Herbert) regime,\cite{Herbert1984} while the transition with $\Lambda$-structures lining up and following each other is termed the K-type regime,\cite{Kachanov1994} in recognition of the pioneering experiments of Klebanoff \textit{et al.}\cite{Klebanoff1962}

\begin{figure*}
\includegraphics{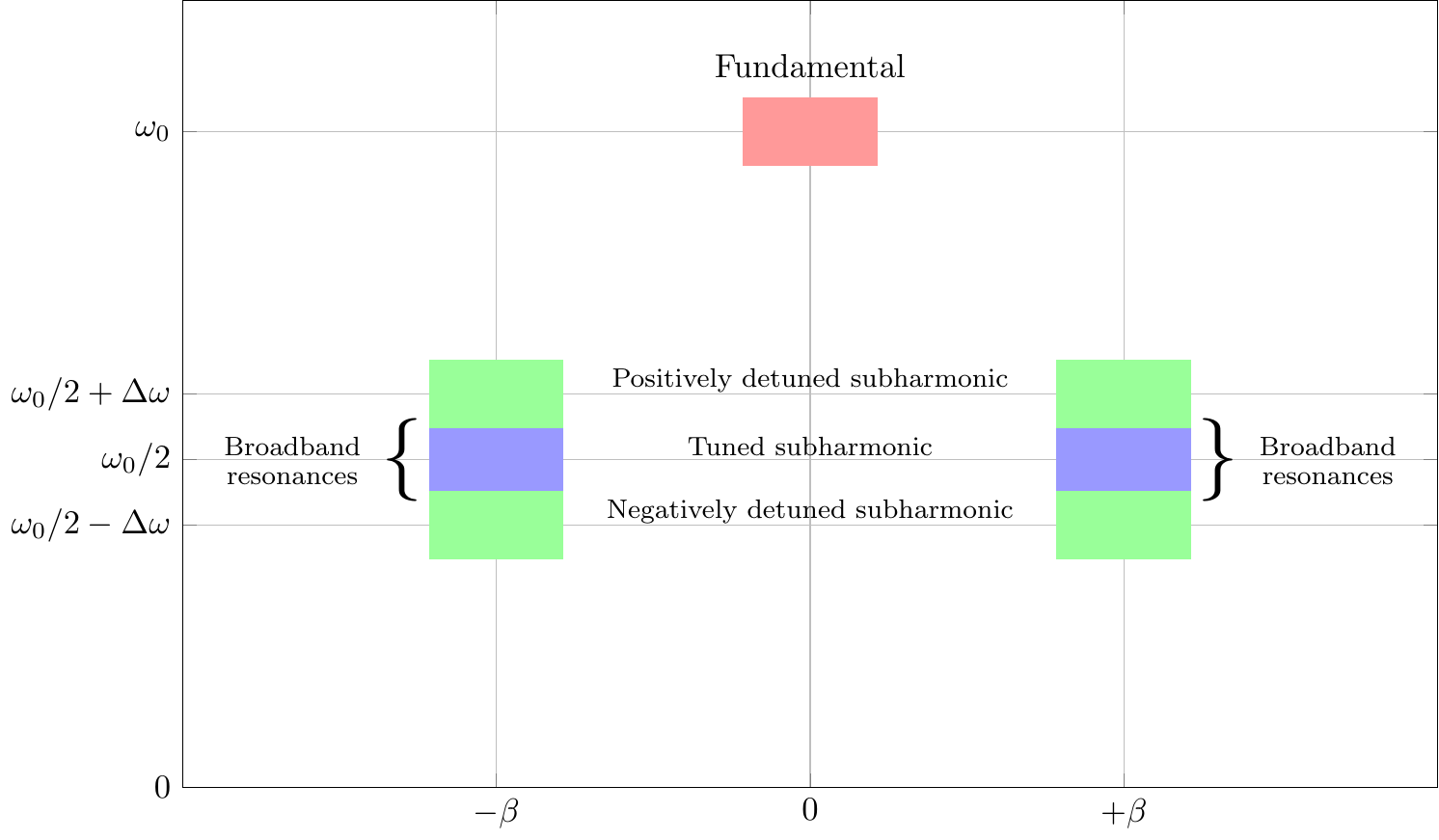}
\caption{\label{fig:schematic-resonance} Schematic of classical resonance mechanisms: the tuned resonant triad of Craik\cite{Craik1971} and the detuned resonances of Herbert.\cite{Herbert1988} Taken together, we have a broad frequency band in which rapid resonant amplification of the subharmonic, catalyzed by the fundamental, may occur.\cite{Borodulin2002c}}
\end{figure*}

Further work showed the detuned resonance to be very wide in the frequency spectrum,\cite{Kachanov1994} with the range $\Delta\omega$ over which resonant amplification could occur being very large and reaching up to half the subharmonic frequency.\cite{Borodulin2002b} And in a paper by W{\"{u}}rz \textit{et al.},\cite{Wurz2012a} the amplification factors across a wide range of frequency and wavenumber detunings were systematically investigated, yielding information on optimal detunings with maximum amplification. For positive frequency detunings in an adverse pressure gradient boundary layer, the amplification factor of the detuned mode could be even greater than the case of tuned resonances!\cite{Wurz2012a} These experiments bring to mind the earlier theoretical work of Wu, Stewart and Cowley\cite{Wu2007} on phase-locked interaction, which is a much less restrictive condition than classical triad resonance. In this phase-locked approach, all 3D disturbances sharing approximately the same phase speed as the 2D mode can be amplified, and a precise subharmonic relation between modes is not required. Wu \textit{et al.}\cite{Wu2007} believe this mechanism to be applicable to both a Blasius boundary layer and a decelerating boundary layer. Such findings suggest that the rather rigid framework of the original resonant triad should be extended into a more flexible mechanism that could accommodate interactions between spectral bands. A schematic of the tuned, detuned and broadband resonance mechanisms in the frequency-spanwise wavenumber ($\omega$-$\beta$) plane is provided in Figure~\ref{fig:schematic-resonance}.

In fact, many disturbances that trigger transition to turbulence in nature possess large spectral widths or bands in space and time - more aptly described as a packet of waves, or a \emph{wavepacket}. Its route or development to turbulence represents in essence a competition for dominance among the many waves present; growing linearly or independently of each other at first, but progressively competing with each other once they attain sufficiently large amplitude in an ever-growing avalanche of nonlinear interactions. This is a crucial motivation for the direct study of wavepackets, where the dominant processes of growth or transition may be studied as they emerge from the complex sum and difference interactions among the numerous modes in the wavepacket. In addition, wavepackets are also created as spatio-temporal perturbations in spatially stable boundary layers through the interaction of multiple stable modes, as studied by Sengupta \textit{et al.}\cite{Sengupta2006,Sengupta2006a} using a Bromwich contour integral method. The works of Breuer, Cohen and Haritonidis,\cite{Breuer1997} Medeiros and Gaster,\cite{Medeiros1999a,Medeiros1999b} Yeo \textit{et al.}\cite{Yeo2010} and others suggest that there is a strong sense of universality in the transition process to turbulence for broadband disturbances that were initiated by pulse-type excitation. 

In particular, Yeo \textit{et al.}\cite{Yeo2010} who modelled the experiments of Cohen \textit{et al.}\cite{Cohen1991} and Breuer \textit{et al.}\cite{Breuer1997} by DNS shows that the high-frequency high-wavenumber spectrally-incoherent wavepackets in the final stage of the experiments rapidly broke down into down-stream-pointing arrowhead shape turbulent spots, which are the basic constituents to a fully turbulent flow. The present work could be regarded as an extension to Yeo \textit{et al.}\cite{Yeo2010} to further probe the make-up or constituents of nonlinear processes/mechanisms that operate and compete within the wavepacket that modelled the experiments of Cohen \textit{et al}.

\section{\label{computational-method} Computational Methodology}
\subsection{\label{subsec:dns-code} DNS code}
In fluid dynamical systems, it is usually not known \textit{a priori} if a disturbance grows either in space or in time or spatio-temporally.\cite{Sengupta2006,Sengupta2006a} To cover a wide range of possibilities, a 3D spatio-temporal DNS is used in this work, and the full details of this DNS code have been published.\cite{Wang2003, Wang2005} Furthermore, a critical study of numerical schemes for transitional boundary layers was presented by Sengupta \textit{et al.}\cite{Sengupta2015b,Sengupta2016} Here, we give an outline of the code for application to our specific needs of simulating a wavepacket evolving in a Blasius boundary layer.

Our DNS code is configured such that it can be run in three different ways, corresponding to the linear perturbation of the Navier-Stokes equations, the nonlinear perturbation, or the full Navier-Stokes equations. Unless otherwise stated, the nonlinear perturbation form is used. The linear perturbation form is used primarily for comparison with the nonlinear results in order to clearly distinguish between the linear and nonlinear phenomena in the flow. Notably, this linear code takes into account the non-parallel, spatially growing boundary layer.

Second-order finite volume spatial discretization and second-order backward Euler temporal discretization is then applied. Our time splitting strategy is a fractional step. This highly efficient technique was developed by Chorin,\cite{Chorin1969} Temam,\cite{Temam1984} and Kim and Moin,\cite{Kim1985} and it has emerged as one of the most popular DNS algorithms in use today. For numerical stability, we use a fully implicit iterative variant of the fractional step method with a pressure correction scheme. Spatially, the DNS code adopts finite volume discretization on a collocated grid system formulated in general curvilinear nonorthogonal coordinates. Collocation of the velocity and pressure data at the same grid points triggers numerical perturbation pressure oscillations that are stabilized by the momentum interpolation method of Rhie and Chow.\cite{Rhie1983} A geometric multigrid procedure\cite{Wesseling2001} is employed to solve the pressure-Poisson problem, with an alternating direction implicit (ADI) solver\cite{Birkhoff1962} in 3D as the smoother and the full approximation storage (FAS) algorithm of Brandt\cite{Brandt1977} used. 

\subsection{\label{subsec:dns-domain} Computational domain and parameters}
The following computational domain and parameters are modeled after the experimental setup of Cohen, Breuer and Haritonidis.\cite{Cohen1991} While the DNS code itself is formulated in general curvilinear coordinates, this paper uses Cartesian coordinates, which are sufficient to investigate our simple domain geometry of a flat plate boundary layer. The streamwise, wall-normal and spanwise Cartesian coordinates are denoted by $x$, $y$ and $z$ respectively. They are non-dimensionalized based on the reference length $\delta_0=2.3182\times 10^{-3}$ m, which is the boundary layer displacement thickness at the disturbance source. If $x^*$, $y^*$ and $z^*$ are dimensional lengths, then $x=x^*/\delta_0$, $y=y^*/\delta_0$ and $z=z^*/\delta_0$. The computational domain is a box with $310\leq x \leq 1510$, $0\leq y \leq 54$, $-173\leq z \leq 173$, and it is meshed with $1186\times 85\times 195$ grid points, with uniform meshing in the $x$ and $z$ direction, and a stretched grid in the $y$ direction to increase grid resolution close to the wall, according to the formula
\begin{equation}
y=\frac{y_{max}\gamma\xi}{\gamma\xi_{max}+y_{max}\left(\xi_{max}-\xi\right)},
\end{equation}
where $\xi$ is the index number of the grid point. Thus, $\xi$ is an integer satisfying $0< \xi \leq \xi_{max}=85$. Similarly, $y$ will be a real number such that $0\leq y \leq y_{max}=54$. $\gamma$ is a stretching parameter that is set to 1.6.

Freestream velocity is $U_\infty = 6.65$ m/s and kinematic viscosity is $\nu = 1.49\times 10^{-5}$ $\text{m}^2/\text{s}$. The disturbance source is located at $x=349.4$, giving rise to a displacement thickness Reynolds number  $\text{Re}=\delta U_\infty /\nu = 1035$, or in terms of momentum thickness Reynolds number $\text{Re}_\theta=\theta U_\infty /\nu = 399$. The Reynolds numbers at the inflow and outflow of the DNS domain are $\text{Re}=975$ ($\text{Re}_\theta=376$) and $\text{Re}=2151$ ($\text{Re}_\theta=830$) respectively. Time is non-dimensionalized as $t=t^*U_\infty /\delta_0$, with $t^*$ measured in seconds. The non-dimensional angular frequency is $\omega = 2\pi f\delta_0 /U_\infty$, where $f$ is the frequency in Hertz. The symbols $u$, $v$ and $w$ represent the streamwise, wall-normal and spanwise perturbation velocities respectively, on a Blasius mean flow profile. These velocities are also non-dimensional; for example, $u\equiv \Delta x^*/\Delta t^*=(\Delta x^*/\delta_0)/(\Delta t^*U_\infty/\delta_0)=(\Delta x^*/ \Delta t^*)/U_\infty$.

The inflow boundary condition is zero perturbation velocity, which is equivalent to a laminar boundary layer inflow. At the outflow boundary, the streamwise second derivative of all velocity components is set to zero; $\partial^2u/\partial x^2=\partial^2v/\partial x^2=\partial^2w/\partial x^2=0$. A buffer domain region\cite{Liu1994} is also implemented just before the outflow boundary, to prevent wave reflections upstream. Periodic boundary conditions are used in the spanwise direction. At the wall, the no slip condition is imposed, while far from the wall, the perturbation velocity is assumed to be zero, corresponding to freestream conditions.

The wavepacket originates from a disturbance source that is a wall-normal, sinusoidal perturbation velocity specified within a circle on the wall. In particular, the initial disturbance is applied to grid points $(x,z)$ on the wall satisfying $\sqrt{(x-x_0)^2+(z-z_0)^2}<R$ where $R=\sqrt{8}$ is the radius of a circular disk centered at $(x_0,z_0)=(349.4,0)$. To impose the disturbance in a spatially smooth manner, the input disturbance function $v(t)$ is multiplied by a two-dimensional Gaussian function such that within the disk:

\begin{equation}
v_{source}(x,z,t)=v(t) \exp\left[-\frac{(x-x_0)^2}{2}-\frac{(z-z_0)^2}{2}\right].
\end{equation}

\begin{equation}
  v(t) = \begin{cases}
      0.0082\sin(0.063t), & 0\leq t \leq \frac{2\pi}{0.063}=99.733,\\
      0, & \text{otherwise}.
  \end{cases} 
\label{eq:initial-disturbance-specific}
\end{equation}

The values of the frequency and amplitude of $v(t)$ were chosen after a detailed study of the combined effects of frequency, amplitude and bandwidth on the nonlinear transition process, as reported in Kang and Yeo.\cite{Kang2013,Kang2015} In short, the initial amplitude was chosen to be small (less than 1\% of freestream velocity) to be within the linear regime. The frequency was chosen such that the peak spectral density of the initial wavepacket corresponds to the lower branch of the neutral stability curve at the actuator location (Reynolds number, $\text{Re}=1035$). This location on the neutral stability curve follows Yeo \textit{et al.},\cite{Yeo2010} which in turn is based on the experiments of Cohen \textit{et al.}\cite{Cohen1991} and Breuer \textit{et al.}\cite{Breuer1997} Medeiros and Gaster\cite{Medeiros1999b} showed that the subharmonic/oblique route to transition is relatively robust with respect to changes in the initial spectral composition of the wavepacket; except for the most drastic spectral cut-off such as discussed by Craik.\cite{Craik2001}

\subsection{\label{subsec:dns-convergence} DNS grid convergence}
To check the adequacy of the grid resolution, the wavepacket simulation was also run at a higher resolution. This was done by first determining the spacing between points of the original $1186\times 85\times 195$ grid: $\Delta x^+=24.554$, $\Delta y^+_{min}=0.4595$ and $\Delta z^+=43.128$. These spacings are given in terms of non-dimensional wall units using the standard formulae $\Delta x^+ = \Delta x^* u_\tau/\nu$ and $u_\tau = \sqrt{\tau_w/\rho}$ where $\Delta x^*$ is the dimensional grid spacing, $u_\tau$ is the friction velocity, $\tau_w$ is the wall shear stress at the source and $\rho$ is the fluid density. Note that $\Delta y^+_{min}$ is the grid spacing in the wall-normal direction just above the wall. The largest grid spacing is in the spanwise $z$-direction, hence this was the target of the most aggressive grid refinement. The refined grid has $1586\times 101\times 391$ points, or $\Delta x^+=18.39375$, $\Delta y^+_{min}=0.3851$ and $\Delta z^+=21.564$. In terms of temporal resolution, the original grid has non-dimensional time step $\Delta t=0.25$ giving rise to a Courant-Friedrichs-Lewy (CFL) number $u \Delta t/\Delta x =0.2472$, while the new grid has $\Delta t=0.2$ to yield CFL $u \Delta t/\Delta x =0.2645$. The results with this refined grid show no significant difference with the lower-resolution results. 

This grid convergence test supplements previous grid convergence or validation studies done with the same DNS code over a very similar flow configuration.\cite{Zhao2007, Yeo2010} Numerical validation of the code had also been performed in Wang, Yeo and Khoo\cite{Wang2005} against the linear and nonlinear results of Fasel \textit{et al.}\cite{Fasel1990b} and Liu \& Liu.\cite{Liu1995}

\section{\label{spectral-method} Spectral Analysis Methodology}
The DNS of the previous section \ref{computational-method} produces data on the time evolution of the velocity field. Spectral analysis using fast Fourier transforms may be performed to identify the spectrum of Fourier modes present at different times. Thereafter, the selection of modes in the frequency-wavenumber domain may be converted back to the space-time domain with an inverse Fourier transform, producing what are commonly known as ``coherent structures''. Alternatively, proper orthogonal decomposition (POD) is a powerful mathematical technique that may be used to find the principal components present in the flow data, although it sometimes produces results that are difficult to understand within the framework of classical theories. Paul and Verma\cite{Paul2016} give a good comparison between POD and Fourier analysis of turbulent signals, discussing the advantages and disadvantages of each technique. In our work, we follow and extend the approach of Sengupta, Swagata and Yogesh\cite{Sengupta2011a} by finding the spectrum of the POD modes, which effectively combines the strengths of both POD and FFT techniques. Coherent structures are first extracted by POD, and then the FFT lends deeper insight by producing a spectrum for each POD structure. In Sengupta \textit{et al.},\cite{Sengupta2011a} a one-dimensional (1D) FFT was used to find the frequency spectrum of the POD modes. This is augmented in our paper to become a 2D FFT, giving the frequency-wavenumber spectrum of the POD modes. Additionally, Sengupta \textit{et al.}\cite{Sengupta2010, Sengupta2011a, Sengupta2011b} pursue a dynamical systems approach to instability, concentrating their discussion on how modes follow or deviate from the Stuart-Landau equation. Our paper pursues the resonance approach instead, drawing connections between the POD modes and the classical theories of Craik and Herbert. While the papers of Sengupta \textit{et al.} use a boundary layer disturbance source that is continually driven, our work focuses on wavepackets which result from a single pulse excitation by the source. The convective nature of the resultant wavepacket disturbance necessitates an extra step to translate the position of the wavepacket before projection onto the POD modes, as detailed in Appendix~\ref{appsubsec:PODmethod-projection}.

\section{\label{sec:FFTmethod}Fourier transform method}
A variety of definitions of Fourier transforms are found in practice, that produce similar results, yet differ numerically primarily because of the use of different normalization factor $1/(2\pi)$ or $1/(\sqrt{2\pi})$. Furthermore, while the transform is often expressed in terms of the angular frequency $\omega$ in the physics and engineering community, mathematicians tend to favor writing in terms of the oscillatory frequency $f$ to avoid breaking the symmetry of the forward-inverse transform pair.\cite{Weisstein2015} In the interest of an accurate description of our work, we set down the definitions of the terms used in our spectral analysis before we present the results.

We first obtain a 2D matrix whose elements $u_{qr}$ are a discrete sampling of the continuous velocity function $u(t,z)$ on a grid with local origin at $(t_l,z_l)$, spatial spacing $\Delta_z$ in the $z$ direction and time step $\Delta_t$ such that
\begin{equation}
u_{qr}=u[t=t_l+(q-1)\Delta_t,z=z_l+(r-1)\Delta_z].
\label{eq:u-matrix2D}
\end{equation}

A discrete Fourier transform can then be applied to this $u_{qr}$ matrix as
\begin{eqnarray}
\hat{u}_{mn}=\frac{1}{N_tN_z}\sum\limits_{q=0}^{N_t-1} \sum\limits_{r=0}^{N_z-1} u_{qr} e^{-\text{i}2\pi (mq/{N_t}+nr/{N_z})}. \label{eq:2d-dft}\\
\qquad m=0,1,\,\ldots,\,N_t-1. \qquad n=0,1,\,\ldots,\,N_z-1.\nonumber
\end{eqnarray}
$N_t$ and $N_z$ are equal to the total number of data points in the $t$ and $z$ dimensions respectively. The spectral coefficients are then found by multiplying each Fourier coefficient $\hat{u}_{mn}$ with its complex conjugate $\overline{\hat{u}_{mn}}$,
\begin{equation}
S_{mn}=|\hat{u}_{mn}|^2=\hat{u}_{mn}\overline{\hat{u}_{mn}}.
\label{eq:spectral-coefficients}
\end{equation}
      
In order to use these spectral coefficients $S_{mn}$ to approximate the spectral density of the wavepacket, we compare Equation~(\ref{eq:2d-dft}) with the definition of spectral density in Equation~(\ref{eq:spectral-density}), which is equivalent to equation (15.29) on page 234 of Newland,\cite{Newland1993}  
\begin{equation}
S(\omega,\beta)=\frac{1}{(2\pi)^2} \int_{-\infty}^{\infty}\text{d}t \int_{-\infty}^{\infty}\text{d}z \enskip R(t,z) e^{-\text{i}(\omega t+\beta z)}.
\label{eq:spectral-density}
\end{equation}
where $R(t,z)$ is the correlation function, $\omega=2\pi f$ is the angular frequency and $\beta$ is the spanwise wavenumber. We then use the approximation
\begin{equation}
S(\omega_m,\beta_n)\approx \frac{L_tL_z}{4\pi^2} S_{mn},
\label{eq:spectral-approximation}
\end{equation}
with $L_t=(N_t-1)\Delta_t$ and $L_z=(N_z-1)\Delta_z$ being the record lengths. Additionally, we have real-numbered $\omega_m=2\pi m/L_t$ and $\beta_n=2\pi n/L_z$. The reasoning used to arrive at this formula is explained in Ref.~\onlinecite{Newland1993}. In exchange for this convenience, our approximation limits our spectral resolution in any dimension to $2\pi/L$, where L is the record length in that dimension.\cite{Kay1981}

Note that while the above explanation used the $u(t,z)$ data to obtain the frequency-spanwise wavenumber $(\omega,\beta)$ spectrum, if we have $u(x,z)$ data instead, it can be used to obtain the streamwise-spanwise wavenumber $(\alpha,\beta)$ spectrum in a similar manner. On a grid with local origin at $(x_l,z_l)$, grid spacing $\Delta_x$ and a total of $N_x$ data points in the $x$-direction, this would yield
\begin{equation}
u_{sr}=u[x=x_l+(s-1)\Delta_x,z=z_l+(r-1)\Delta_z],
\label{eq:u-matrix2D-xz}
\end{equation}
\begin{eqnarray}
\hat{u}_{pn}=\frac{1}{N_xN_z}\sum\limits_{s=0}^{N_x-1} \sum\limits_{r=0}^{N_z-1} u_{sr} e^{-\text{i}2\pi (ps/{N_x}+nr/{N_z})}. \label{eq:2d-dft-xz}\\
\qquad p=0,1,\,\ldots,\,N_x-1. \qquad n=0,1,\,\ldots,\,N_z-1.\nonumber
\end{eqnarray}
We then approximate the $S(\alpha,\beta)$ spectral density as
\begin{equation}
S(\alpha_p,\beta_n)\approx \frac{L_xL_z}{4\pi^2} S_{pn}=\frac{L_xL_z}{4\pi^2} |\hat{u}_{pn}|^2,
\label{eq:spectral-approximation-xz}
\end{equation}
with $L_x=(N_x-1)\Delta_x$, $L_z=(N_z-1)\Delta_z$, $\alpha_p=2\pi p/L_x$ and $\beta_n=2\pi n/L_z$.
 
\section{\label{sec:PODmethod}Proper orthogonal decomposition (POD) method}
Given a square-integrable, complex-valued function $u(x)$ belonging to the linear, infinite-dimensional Hilbert space $L^2$ on domain $\Omega$ with inner product
\begin{equation}
(f,g)=\int_{\Omega} f(x)\overline{g(x)}\text{d}x,
\end{equation}
the proper orthogonal decomposition (POD) finds a basis $\lbrace \phi_j(x)\rbrace_{j=1}^{\infty}$ that is optimal in the sense that the average squared error between $u$ and its projection onto this basis is minimized.\cite{Holmes2012} (The overline $\overline{g(x)}$ represents the complex conjugate of $g(x)$.) In this basis, for a finite-dimensional case, $u(x)$ may be expressed as a linear combination of eigenfunctions $\phi_j$ with coefficients $a_j$ such that
\begin{equation}
u_N(x)=\sum_{j=1}^{N} a_j \phi_j(x),
\label{eq:POD-finite-dimensional}
\end{equation}
and the POD becomes an optimal decomposition in the sense that these first $N$ POD basis functions capture more energy on average than the first $N$ functions of any other basis. In an intuitive sense, the decomposition captures the bulk of energetic activities within the disturbance in the least number of independent modes. Hence it is an ideal and neutral tool for isolating and extracting $L^2$-dominant or energetic events or processes subsumed within large data streams or sets.
It has been proven that the POD can be obtained through a singular value decomposition (SVD).\cite{Holmes2012} More details about our computational implementation of the POD and SVD have been placed in Appendix~\ref{appsec:POD-appendix} of this paper.

As just described, the POD requires a definition of the average squared error between $u$ and its projection onto the basis $\lbrace \phi_j(x)\rbrace_{j=1}^{\infty}$. In the context of our wavepacket studies, this is an average in space and/or time. Because the wavepacket laminar turbulent transition is essentially a non-stationary process, to obtain a meaningful average, we view the data through a sliding window, within which the process can be assumed to be quasi-stationary. 

\begin{figure*}
\includegraphics{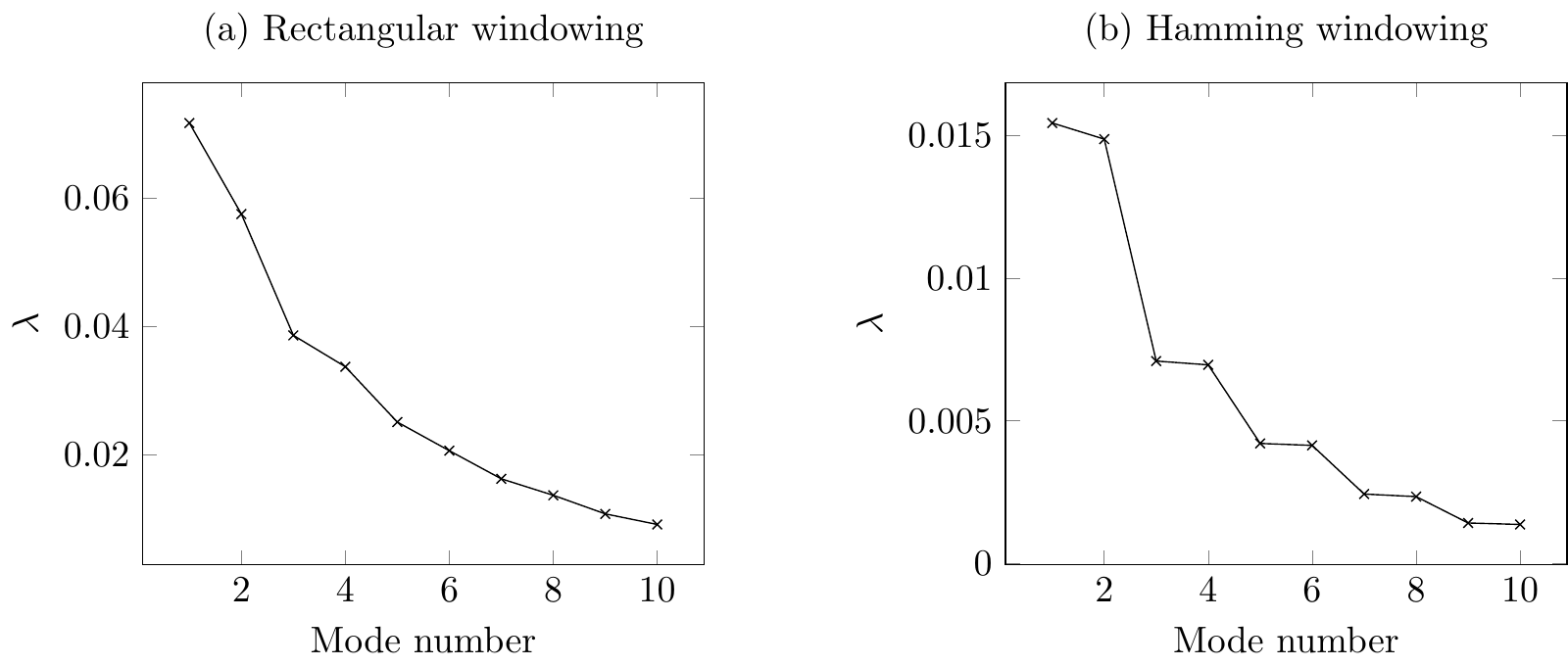}
\caption{\label{fig:POD-spectrum-window-comparison}POD empirical eigenvalue spectrum for the wavepacket $u(t,z)$ velocity at $x=906$, comparing rectangular windowing in (a) with Hamming windowing in (b). The windows have dimensions $\Delta t=1344$ and $\Delta x=302$.}
\end{figure*}

Guided by the review paper of Harris,\cite{Harris1978} tests were conducted using both rectangular and Hamming windows, and it can be seen from Figure~\ref{fig:POD-spectrum-window-comparison} that the Hamming window produced a sharper and more distinct pattern of POD eigenvalues. (It is known that our POD eigenvalues and eigenfunctions should occur in pairs. Such pairing of modes arise because the space-time symmetry of a traveling wave leads to a degenerate POD eigenproblem, such as in a parallel flow.\cite{Rempfer1994b} But since our Blasius boundary layer is slowly spatially growing, our POD eigenproblem is only near degenerate and the pairs of eigenvalues are not exactly equal.\cite{Rempfer1994a}) Such Hamming windowing may be thought of as a weighted average, with maximum weight being assigned to the center of the window. In particular, low weights at the downstream edge of the window are important to reduce the influence of high-amplitude, late-stage structures that have newly formed in the wavepacket, for these are not representative of the state of the wavepacket throughout the rest of the window. Thus, before POD, Hamming windowing was applied to the wavepacket data set in the time $t$ and streamwise $x$ dimensions, while a rectangular window was sufficient in the spanwise $z$ direction because of the spanwise periodic boundary conditions of the DNS computational domain.

\section{\label{results} Results}

\begin{figure*}
\includegraphics{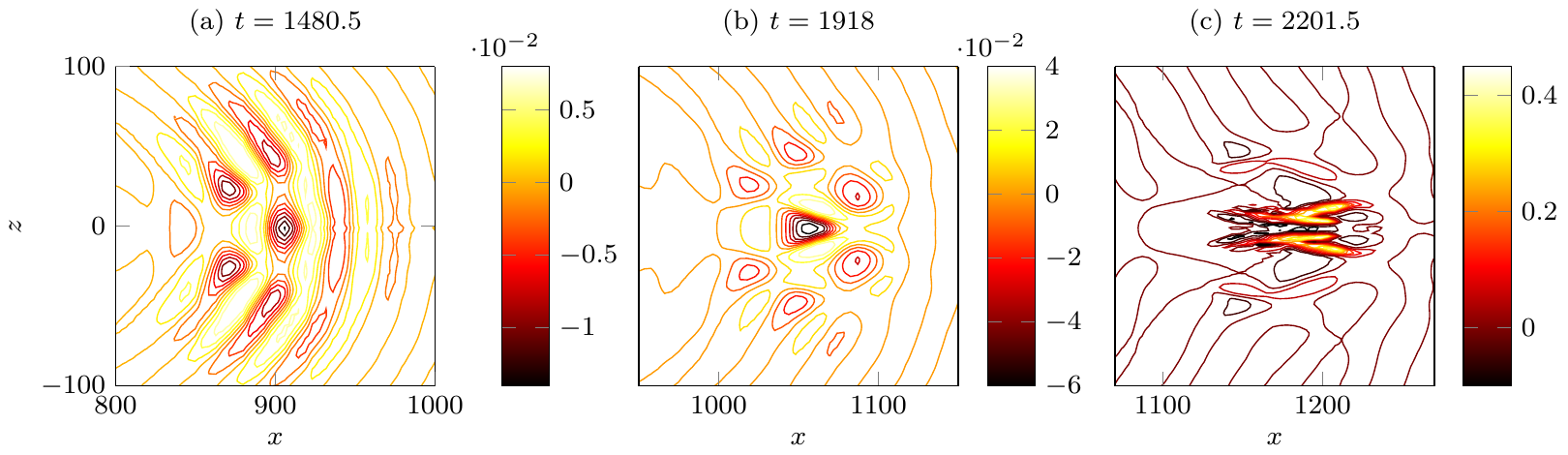}
\caption{\label{fig:single-cycle-u-iso-slice-X906}Streamwise perturbation $u$-velocity contours of the DNS wavepacket in the $x$-$z$ plane with $y^*/\delta=0.6$ at (a) $t=1480.5$, (b) $t=1918$ and (c) $t=2201.5$.}
\end{figure*}

The DNS wavepacket evolution as it is convected downstream by the flow is displayed at three snapshots in time in Figure~\ref{fig:single-cycle-u-iso-slice-X906}. It can be seen that at $t=1480.5$, the wavepacket is in a weakly nonlinear stage of development, with some spanwise variation across the central crescent ridges. Moving on to $t=1918$, the center ridge has become a triangular-shaped depression that develops into a horseshow vortex by $t=2201.5$. At this point, there are concentrated pockets of highly sheared flow in the core of the wavepacket, indicative of incipient turbulence. For the purposes of this study, the subsequent focus will be on the weakly nonlinear stages of transition. The interested reader is invited to refer to Yeo \textit{et al.}\cite{Yeo2010} for a more detailed exposition of the full transition process undergone by this wavepacket.

\begin{figure*}
\includegraphics{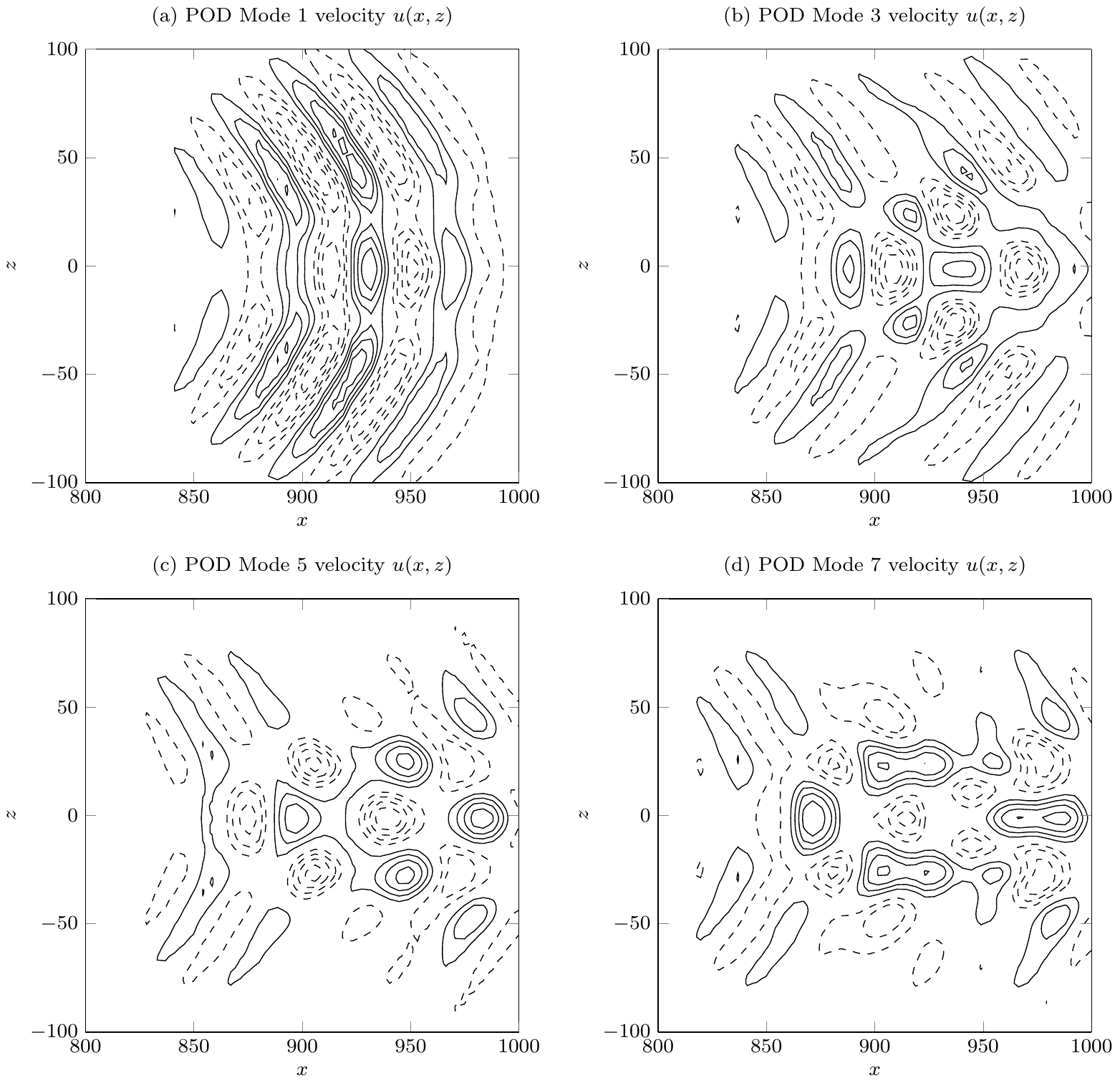}
\caption{\label{fig:single-cycle-pod-X906}Normalized perturbation velocity contours $(u/{|u|}_{\mathrm{max}})$ of the wavepacket modes obtained from POD in the $x$-$z$ plane centered at $x=906$. Solid contour lines denote positive velocity, and dashed lines show negative velocity, with uniformly spaced contour levels $\pm\lbrace0.2,0.4,0.6,\dotsc\rbrace$. The $u=0$ contour is not drawn.}
\end{figure*}

\begin{figure*}
\includegraphics{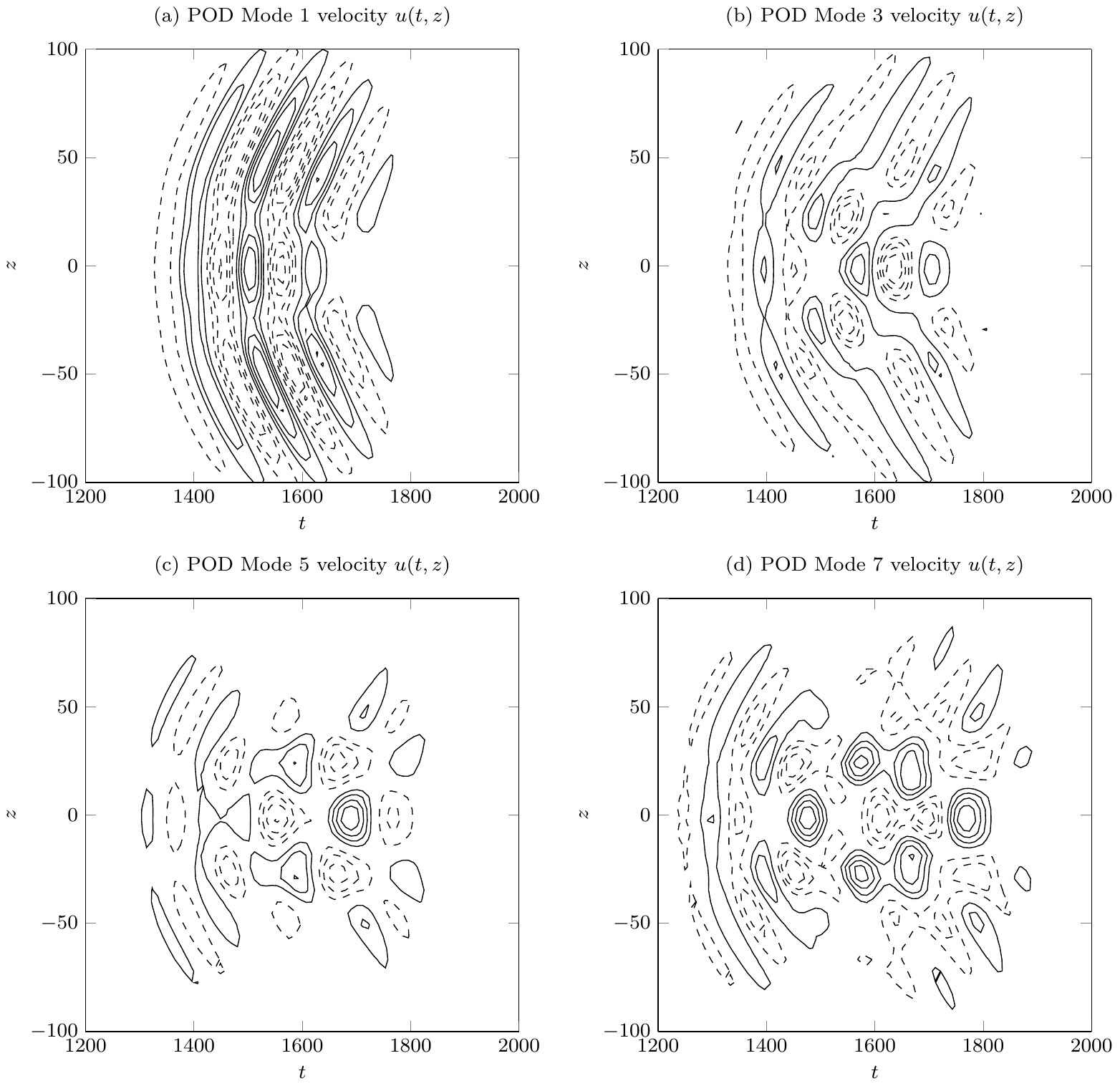}
\caption{\label{fig:single-cycle-pod-X906-spacetime} Normalized perturbation velocity contours $(u/{|u|}_{\mathrm{max}})$ of the wavepacket modes obtained from proper orthogonal decomposition (POD) in the $t$-$z$ plane centered at $t=1480.5$. Solid contour lines denote positive velocity, and dashed lines show negative velocity, with uniformly spaced contour levels $\pm\lbrace0.2,0.4,0.6,\dotsc\rbrace$.  The $u=0$ contour is not drawn.}
\end{figure*}

The proper orthogonal decomposition (POD) of the wavepacket at $t=1480.5$ is shown in Figure~\ref{fig:single-cycle-pod-X906}, obtained according to the methodology described in Section~\ref{sec:FFTmethod}. (To be precise, Figures~\ref{fig:single-cycle-pod-X906} and \ref{fig:single-cycle-pod-X906-spacetime} depict the POD eigenmodes of the disturbance wavepacket generated in Section~\ref{subsec:dns-domain} calculated with the SVD scheme of Appendix~\ref{appsubsec:PODmethod-svd}. Figure~\ref{fig:single-cycle-pod-X906} is the result of a POD on $u(x,z)$ data given by matrix $\mathbf{X}_A$ in Equation~(\ref{eq:data-matrix}), while Figure~\ref{fig:single-cycle-pod-X906-spacetime} is the result of a POD on $u(t,z)$ data represented by matrix $\mathbf{X}_B$ in Equation~(\ref{eq:data-matrix-tz}).) Only the odd-numbered modes are shown, numbered according to the magnitude of their eigenvalues, with mode 1 being the largest. The even-numbered modes are not shown because the POD modes occur in almost-identical pairs, as described in Section~\ref{sec:PODmethod}.

The first POD mode in Figure~\ref{fig:single-cycle-pod-X906}(a) is closest to a 2D fundamental mode, with wave crests and troughs roughly perpendicular to the streamwise $x$ direction. Nevertheless, the POD Mode 1 exhibits some 3D character in the curved crescent shape of the waves, and spanwise modulation of the center region. The third POD mode, shown in Figure~\ref{fig:single-cycle-pod-X906}(b) has a distinctive arrowhead-shaped structure at its front, bringing to mind the well-known $\Lambda$-vortex.\cite{Pierce2013} 

POD Mode 5 in Figure~\ref{fig:single-cycle-pod-X906}(c) appears to be a mix of the fundamental and subharmonic modes, with the subharmonic mode giving rise to a checkerboard-like pattern in the core of the wavepacket due to periodicity in both the $x$ and $z$-directions. POD Mode 7 in Figure~\ref{fig:single-cycle-pod-X906}(d) has parts that are elongated in the streamwise $x$-direction, partially representing the boundary layer streaks or Klebanoff modes. From these mode shapes, it can be seen that a key feature of the POD is that it took into account the finite spatial extent of the wavepacket at the modal level, allowing POD to give concise expression to the compact nature of the wavepacket.

The POD of the same Hamming windowed wavepacket is shown in $t$-$z$ planes in Figure~\ref{fig:single-cycle-pod-X906-spacetime}. These $t$-$z$ planes are centered at time $t=1480.5$, at which the wavepacket's spatial center is around $x=906$. They are broadly similar to the POD modes in the $x$-$z$ plane in Figure~\ref{fig:single-cycle-pod-X906}, but represented in terms of the convective translation of these POD modes in time past $x=906$. The mutual consistency of the POD eigenmodes in the $x$-$z$ and $t$-$z$ planes reflects the semi-permanent character or persistence of these physical events in space and time (as opposed to purely transitory events), which facilitate their study. 

\subsection{\label{subsec:coherent-structures-broadband-hybrid-pod-fft} Hybrid POD-FFT spectrum}
\subsubsection{\label{subsubsec:coherent-structures-broadband-hybrid-pod-fft-xz} Hybrid POD-FFT spectrum for x-z plane data}

\begin{figure*}
\includegraphics{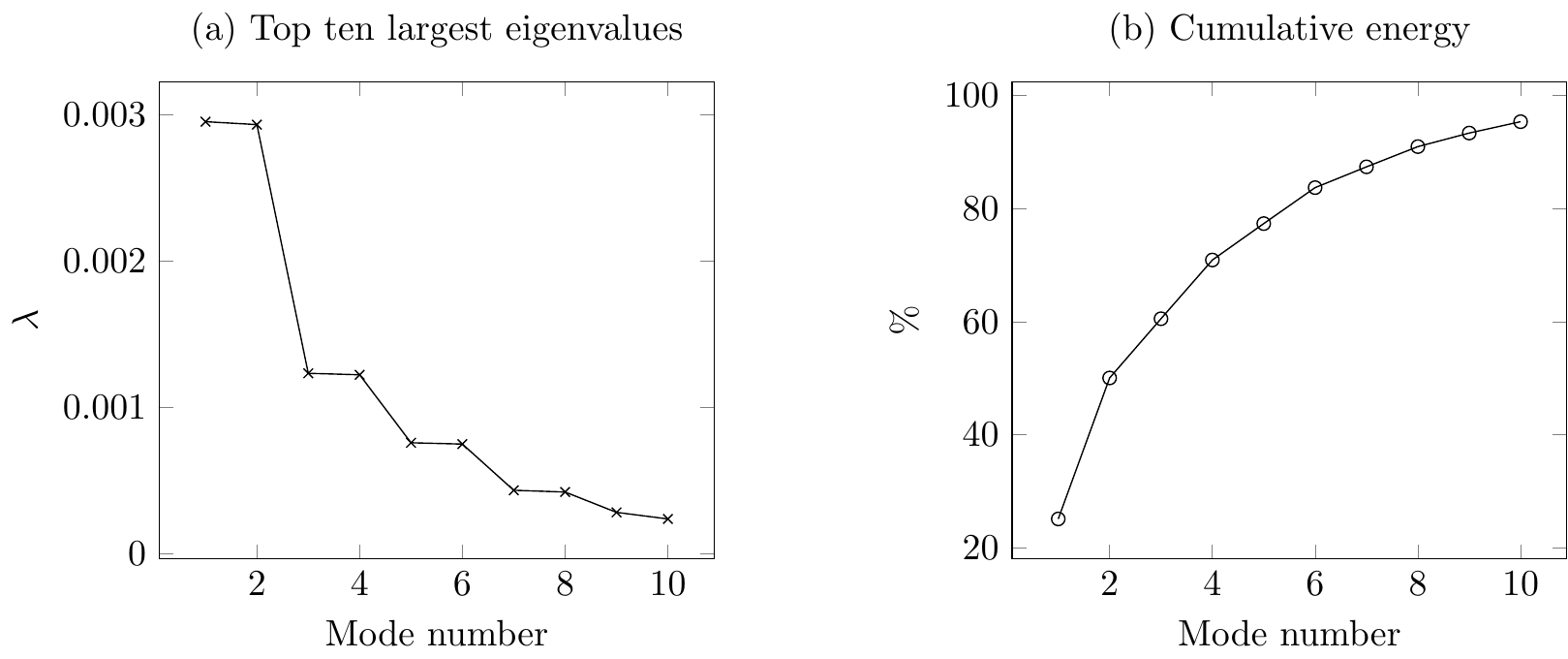}
\caption{\label{fig:POD-spectrum}POD empirical eigenvalue spectrum and cumulative energy for the $u(x,z)$ wavepacket centered at $x=906$.}
\end{figure*}

More interesting insight into these POD results may be obtained by finding the spectral density of each POD mode, and we term this the hybrid POD-FFT. The kinetic energy of the POD modes are represented by their empirical eigenvalues $\lambda_j$ (see Section~\ref{sec:PODmethod} for details), shown in Figure~\ref{fig:POD-spectrum}(a) for the wavepacket POD in $x$-$z$ planes. The contribution of a particular POD mode to the total kinetic energy of the wavepacket can be expressed as $\lambda_j/\sum_{j=1}^{N} \lambda_j$, where $N$ is the total number of POD modes. Therefore, the energy captured by the first $i$ modes can be expressed as\cite{Liang2015} 
\begin{equation}
\text{\% total energy}=\frac{\sum_{j=1}^{i} \lambda_j}{\sum_{j=1}^{N} \lambda_j}\times100\%. 
\end{equation}
This equation is used to plot the cumulative energy shown in Figure~\ref{fig:POD-spectrum}(b). The first two POD modes are almost identical, and they are found to capture 50.04\% of the total kinetic energy of the wavepacket $u$-velocity component. The first ten POD modes cumulatively carry 95.37\% of the total energy. Thus, the POD can be seen as a method for filtering away ``noise'' and focusing the spectral analysis on the most important features of the flow.\cite{Ichihashi2010} 

\begin{figure*}
\includegraphics{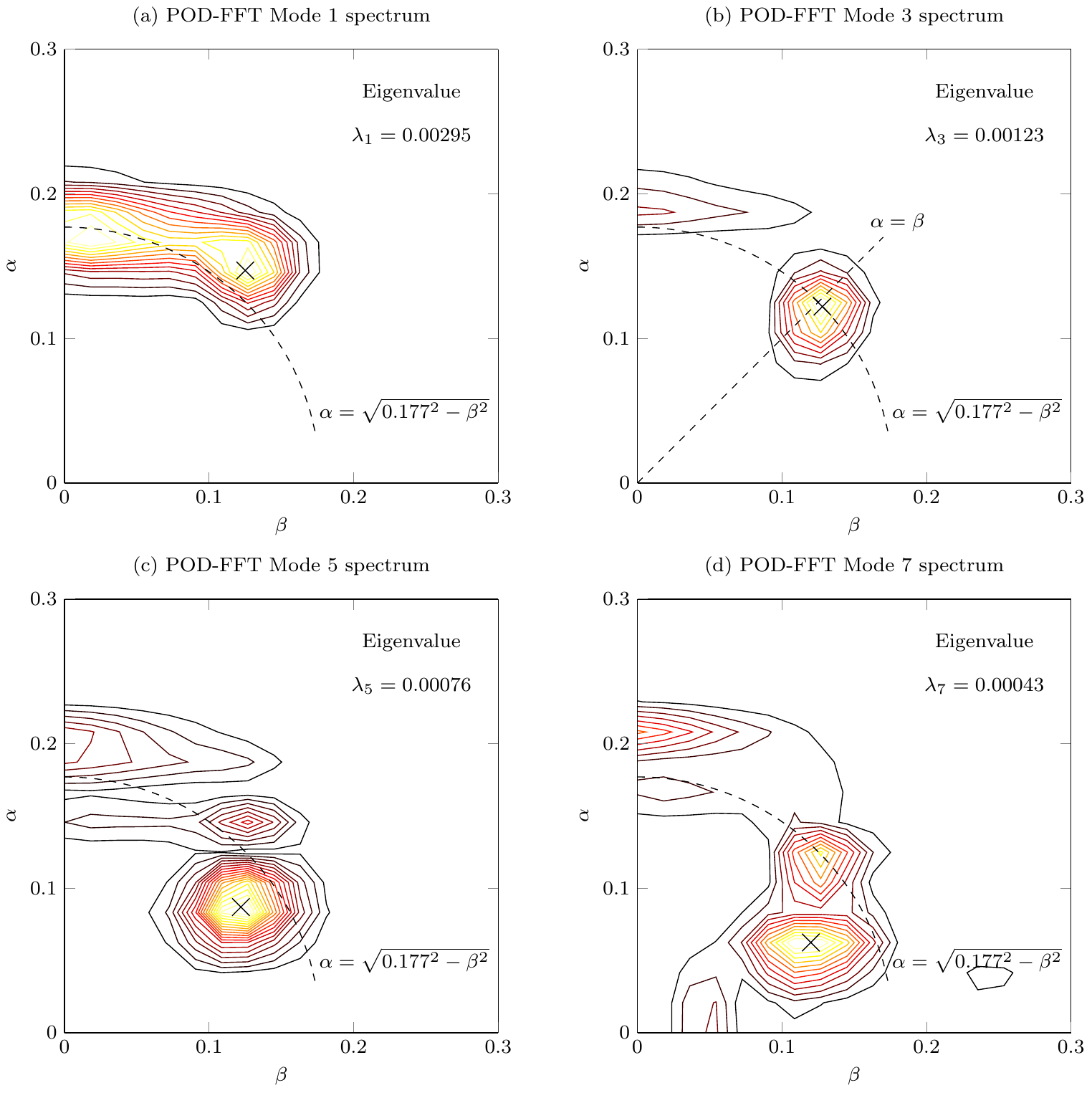}
\caption{\label{fig:single-cycle-pod-X906-fft} Hybrid POD-FFT results, showing the spectral density of the wavepacket POD modes $u(x,z)$ in Figure~\ref{fig:single-cycle-pod-X906}. The dominant oblique mode in each spectrum is marked with ``$\times$''.}
\end{figure*}

Figure~\ref{fig:single-cycle-pod-X906-fft} presents the $(\beta,\alpha)$ spectral density of the POD modes of Figure~\ref{fig:single-cycle-pod-X906}. Since the POD eigenvalue already characterizes the magnitude/energy of the individual POD mode, the contours of spectral density may not be accompanied by a legend showing their absolute magnitude. Instead, the emphasis in Figure~\ref{fig:single-cycle-pod-X906-fft} will be to show the \emph{relative} distribution or concentration of energy among the leading Fourier spectral components within the POD modes.

Figure~\ref{fig:single-cycle-pod-X906-fft}(a) shows the spectral density of the first POD mode. Scrutiny of this figure shows that it consists of two local maxima: one at $(\beta,\alpha)=(0,0.17)$ and another at $(\beta,\alpha)=(0.125,0.145)$. The first local maximum $(\beta,\alpha)=(0,0.17)$ may be associated with the crescent-shaped ripples in the wavepacket. This is because it occupies a band of almost constant streamwise wavenumber $\alpha$, and a range of spanwise wavenumber from $\beta=0$ to about $\beta=0.09$. Since the wave propagation direction is given by $\mathbf{k}/|\mathbf{k}|$, where $\mathbf{k}$ is the wavenumber vector, it follows that the propagation angle of the waves in this mode relative to the streamwise $x$-axis varies from $0^\circ$ to $\arctan(\beta/\alpha)=\arctan(0.09/0.17)=28^\circ$. Referring back to Figure~\ref{fig:single-cycle-pod-X906}(a), we indeed find that the propagation angle of the waves in POD Mode 1 vary smoothly in an arc from $0^\circ$ at the centerline $z=0$ to around $\pm30^\circ$ at the sides of the wavepacket crescent. The second local maximum in the Mode 1 spectrum at $(\beta,\alpha)=(0.125,0.145)$ has a spanwise wavenumber $\beta$ that produces a wave of spanwise wavelength $2\pi/\beta=2\pi/0.125\approx50$, which seems close to the width of the spanwise modulation of Mode 1 in Figure~\ref{fig:single-cycle-pod-X906}(a). 

Proceeding now to the POD-FFT Mode 3 in Figure~\ref{fig:single-cycle-pod-X906-fft}(b), we find that it contains a single strong mode at $(\beta,\alpha)=(0.128,0.122)$. This means that it almost lies along the line $\alpha=\beta$, and the wave has propagation angle $\arctan(\beta/\alpha)\approx45^\circ$. Returning to the associated velocity contours in Figure~\ref{fig:single-cycle-pod-X906}(b), we see the presence of a large $\Lambda$-vortex. The two sides/legs of this ``V'' shape are angled at almost precisely $\pm45^\circ$ to the streamwise direction of the flow, supporting a conclusion that this $(\beta,\alpha)=(0.128,0.122)$ mode is indeed a $\Lambda$-vortex. The spectrum of this mode also follows the Squire transformation line $\alpha=\sqrt{0.177^2-\beta^2}$ drawn dashed in Figure~\ref{fig:single-cycle-pod-X906-fft}(b) for Squire wavenumber $\widetilde{\alpha}=0.177$. At this juncture, we should recall that the Squire transformation means that every 3D wave eigenvalue problem for parallel flow can be reduced to an equivalent 2D ($\beta=0$) eigenvalue problem.\cite{Squire1933} This transformation is achieved by a rotation of the coordinate reference frame into the wave propagation direction and application of velocity scaling. The equivalent 2D wavenumber $\widetilde{\alpha}$ is related to the 3D wavenumbers $(\beta,\alpha)$ by $\widetilde{\alpha}=\sqrt{\alpha^2+\beta^2}$. Hence, the 3D wave $(\beta,\alpha)=(0.128,0.122)$ is equivalent to a 2D wave $(\widetilde{\beta},\widetilde{\alpha})=(0,\sqrt{0.128^2+0.122^2})=(0,0.177)$. The dashed line $\alpha=\sqrt{0.177^2-\beta^2}$ therefore represents the entire family of 3D waves that are equivalent to the 2D $\widetilde{\alpha}=0.177$ wave. This $\alpha=\sqrt{0.177^2-\beta^2}$ line is also drawn in Figure~\ref{fig:single-cycle-pod-X906-fft}(a), and it can be seen that it passes almost directly through the 2D energy peak of Mode 1 at $(\beta,\alpha)=(0,0.17)$.

The hybrid POD-FFT Mode 5 of Figure~\ref{fig:single-cycle-pod-X906-fft}(c) has the dominant mode $(\beta,\alpha)=(0.122,0.087)$. This $\alpha=0.087$ value is almost exactly half that of POD Mode 1's 2D component at $\alpha=0.17$, meaning that it is forming an approximate Craik triad resonance with it. Contrastingly, POD-FFT Mode 7 in Figure~\ref{fig:single-cycle-pod-X906-fft}(d) has two oblique modes at $\beta=0.12$ with positive and negative $\alpha$-detunings from the resonant triad, at $\alpha=0.095+0.028=0.123$ and $\alpha=0.095-0.032=0.063$, which is consistent with the detuned mode pair of the Herbert secondary instability theory.\cite{Herbert1988} The positively detuned mode in this pair passes through the same $\alpha=\sqrt{0.177^2-\beta^2}$ line as Mode 3 and the 2D energy peak of Mode 1, showing that all these modes are part of the same Squire mode family and can satisfy the relaxed matching conditions of Wu \textit{et al.}\cite{Wu2007} for approximate phase-locked interaction. These results coupled with the magnitude information of the POD modes clearly reflect the relative contribution of the various mechanisms to the development of the wavepacket.

\subsubsection{\label{subsubsec:coherent-structures-broadband-hybrid-pod-fft-tz} Hybrid POD-FFT spectrum for t-z plane data}

\begin{figure*}
\includegraphics{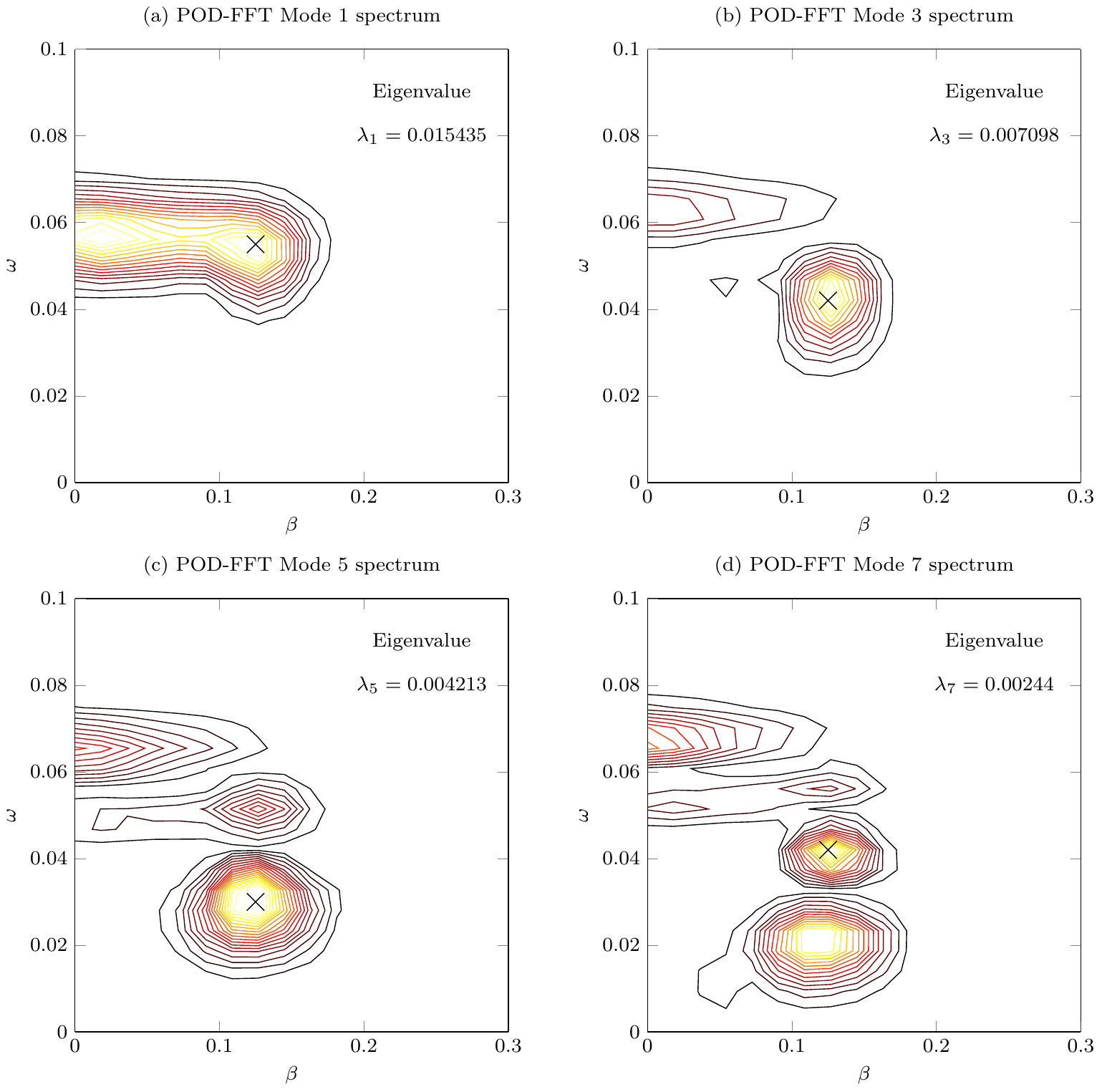}
\caption{\label{fig:single-cycle-pod-X906-fft-spacetime} Hybrid POD-FFT results, showing the spectral density of the wavepacket POD modes $u(t,z)$ in Figure~\ref{fig:single-cycle-pod-X906-spacetime}. The dominant oblique mode in each spectrum is marked with ``$\times$''.}
\end{figure*}

Figure~\ref{fig:single-cycle-pod-X906-fft-spacetime} displays the $(\beta,\omega)$ spectral density of the POD modes of Figure~\ref{fig:single-cycle-pod-X906-spacetime}. Figure~\ref{fig:single-cycle-pod-X906-fft-spacetime}(a) shows that the most energetic POD mode 1 occupies a single band of spectral energy around the dominant initial frequency of the wavepacket, $\omega_0=0.056$. It has two local maxima at $\beta\approx0$ and $\beta\approx0.125$, and so has both 2D and 3D character. The latter peak, with $\omega$ and $\alpha$ close to those of the fundamental 2D wave, may be related to Herbert's fundamental resonance.

The spectrum of the POD modes 3, 5 and 7 in Figures~\ref{fig:single-cycle-pod-X906-fft-spacetime}(b), \ref{fig:single-cycle-pod-X906-fft-spacetime}(c) and \ref{fig:single-cycle-pod-X906-fft-spacetime}(d) all have peak spectral density along the $\beta\approx0.125$ line. The key difference is in the frequency of these peaks, for Mode 3 in Figure~\ref{fig:single-cycle-pod-X906-fft-spacetime}(b), the peak at $\omega_0/2+\Delta\omega=0.056/2+0.014\approx0.042$ seems to be a positively detuned subharmonic mode, with detuning $\Delta\omega=+0.014$ relative to the subharmonic frequency of $\omega_0/2=0.028$ (half the fundamental frequency). It should be highlighted that this detuned mode is not in the sense of the secondary instability of Herbert,\cite{Herbert1988} because it is not accompanied by a complex conjugate pair at a negative frequency detuning, which is required by the theory. The detuned mode is spatially a $\Lambda$-type vortex, as already noted earlier in Section~\ref{subsubsec:coherent-structures-broadband-hybrid-pod-fft-xz}, with a ``V'' shape structure in both the $x$-$z$ and $t$-$z$ domains in Figures~\ref{fig:single-cycle-pod-X906}(b) and \ref{fig:single-cycle-pod-X906-spacetime}(b) respectively.

Mode 5 in Figure~\ref{fig:single-cycle-pod-X906-fft-spacetime}(c) is the closest to an exact subharmonic in a Craik triad,\cite{Craik1971} occupying $\omega=0.030$. This is very close to half of the dominant fundamental 2D frequency in Figure~\ref{fig:single-cycle-pod-X906-fft-spacetime}(a) since $\omega_0/2=0.056/2=0.028$. We also note that the Mode 5 spectrum depicts a 2D $\beta=0$ mode at $\omega=0.065$. The dominant peak of Mode 5 might thus be resonating with this 2D mode and/or the 2D fundamental mode $\omega_0=0.056$ (in POD Mode 1). These frequency coincidences and the satisfaction of the wavenumber criterion noted earlier for Figure~\ref{fig:single-cycle-pod-X906-fft}(c) point unequivocally to the Craik-triad origin of the dominant spectral mode in POD Mode 5. Mode 7 in Figure~\ref{fig:single-cycle-pod-X906-fft-spacetime}(d) seems to have to have a pair of subharmonic modes at positive and negative frequency detunings from $\omega_0/2=0.028$, with the positively detuned peak being slightly stronger than its negatively detuned counterpart. We also recall that in Section~\ref{subsubsec:coherent-structures-broadband-hybrid-pod-fft-xz}, we found that Mode 7 has a positive and negative detuning in its $\alpha$-wavenumber. Since the complex conjugate components in the secondary disturbance theory of Herbert\cite{Herbert1988} predict frequency and wavenumber detuning to be conjoint with each other, there is strong evidence that Mode 7 is indeed the combination resonance of Herbert.\cite{Herbert1988} Nevertheless, it is also a relatively weak mechanism in the wavepacket as a whole, containing about a third of the energy of Mode 3 and half of the energy of Mode 5. 

Considering now the phase speed of the modes, $c=\omega/\alpha$, we combine the results of Figure~\ref{fig:single-cycle-pod-X906-fft} and Figure~\ref{fig:single-cycle-pod-X906-fft-spacetime} to obtain the phase speed of Mode 1's 2D component: $\omega/\alpha\approx 0.056/0.17=0.329$ and the phase speeds of the dominant 3D oblique component of Mode 3: $\omega/\alpha\approx 0.042/0.122=0.344$, Mode 5: $\omega/\alpha\approx 0.030/0.087=0.345$ and Mode 7: $\omega/\alpha\approx 0.021/0.063=0.333$. We find that all these phase speeds are within 3\% of the Blasius base flow velocity at this location in the boundary layer, that is $U=0.340$ at $x=906$, $y^*/\delta=0.6$, showing that the modes are in the vicinity of the $U=c$ critical layer. Waves that are synchronized or phase-locked in such a manner tend to experience strong nonlinear interactions. In this regard, we may recall that the dominant positively-detuned 3D spectral mode in POD Mode 3 has a Squire wavenumber that is close to that of the fundamental 2D mode. According to Wu \textit{et al.}, this wavenumber condition and an optimal small mismatch in phase speed may produce nonlinear phase-locked growth of the mode that is larger (at least initially) than that of tuned resonance.\cite{Wu2007}

Taken together, the POD Modes 3, 5 and 7 show that the wavepacket's subharmonic mode, which appears to stay above half the fundamental frequency in the experiments of Cohen, Breuer and Haritonidis\cite{Cohen1991} and Medeiros and Gaster,\cite{Medeiros1999a,Medeiros1999b} can be separated via POD into at least three distinct modes that occupy $\omega_0/2+\Delta\omega$, $\omega_0/2$ and $\omega_0/2-\Delta\omega$. The energy hierarchy of the modes is clear: the positively detuned mode is strongest, followed by the tuned mode and lastly the negatively detuned mode. Furthermore, the POD suggests that the wavepacket is experiencing multiple wave resonance and growth mechanisms concurrently, with linear amplification, the Craik resonance triad and Herbert's secondary instability all contributing to the growth of the wavepacket, although it would appear that Herbert's parametric mechanism is not the dominant physical process, because Mode 7 \& 8 contribute just 7.51\% of the total kinetic energy of the wavepacket, whereas Mode 3 \& 4 (the dominant subharmonic POD mode) contain around 20.9\% from Figure~\ref{fig:POD-spectrum}(b).

Several possibilities may explain the dominance of the positively-detuned mode over the other subharmonic modes - it might have been seeded at a higher amplitude in the initial spectrum or it might have enjoyed strong early linear growth due to its spectral proximity to the linearly-dominant Mode 1. Another plausible explanation is that positive frequency detunings could experience amplification factors even greater than tuned resonances, fitting into the framework of Borodulin \textit{et al.}\cite{Borodulin2002b} and W{\"{u}}rz \textit{et al.},\cite{Wurz2012a} who describe asymmetry between the amplification of positively and negatively detuned quasi-subharmonic modes, with larger amplification of the positively detuned modes. However, these published results are for an adverse pressure gradient boundary layer, and our results appear to be a first of their kind for a Blasius boundary layer. We would like to stress that the above factors (higher initial amplitude and higher amplification rates) could be jointly contributing to the strength of the positively detuned subharmonic.

\begin{figure*}
\includegraphics{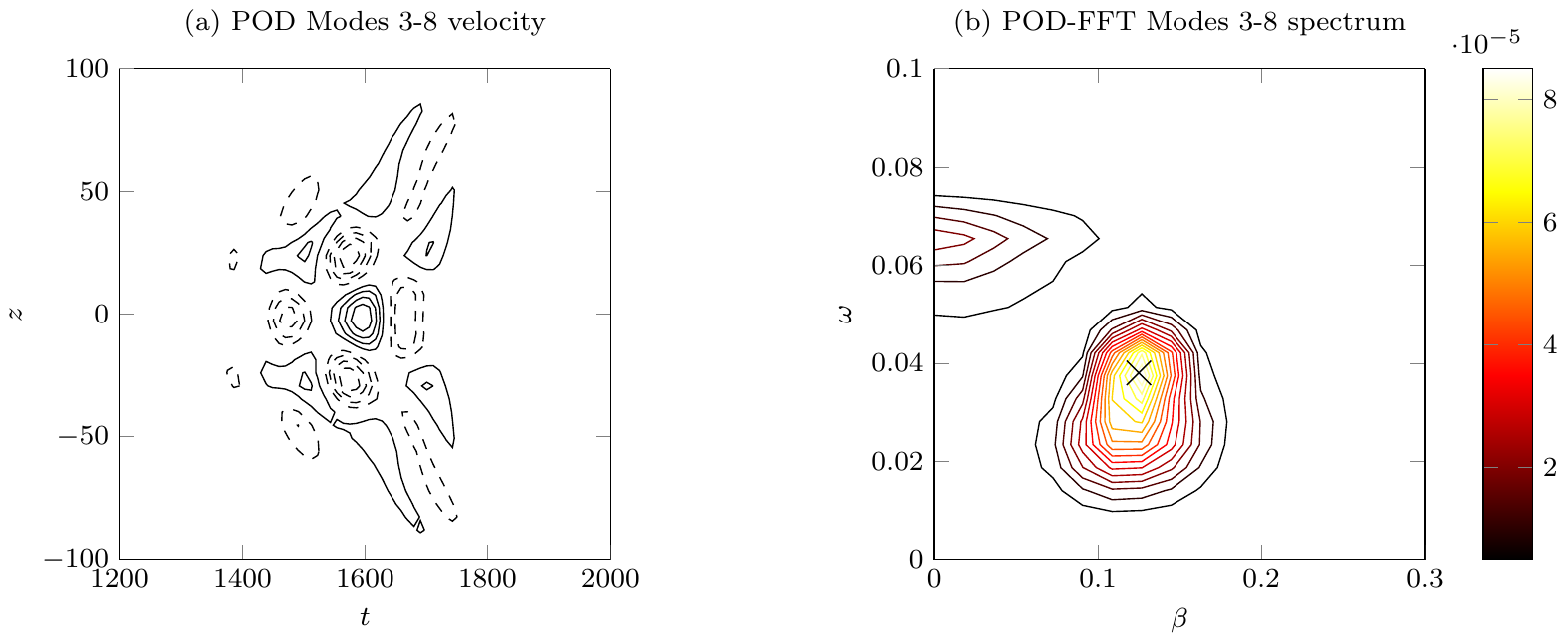}
\caption{\label{fig:single-cycle-pod-X906-fft-spacetime-combi} Combination of wavepacket POD modes 3 until 8, and its spectral density at $x=906$. The dominant oblique mode is marked ``$\times$''.}
\end{figure*}

\begin{figure*}
\includegraphics{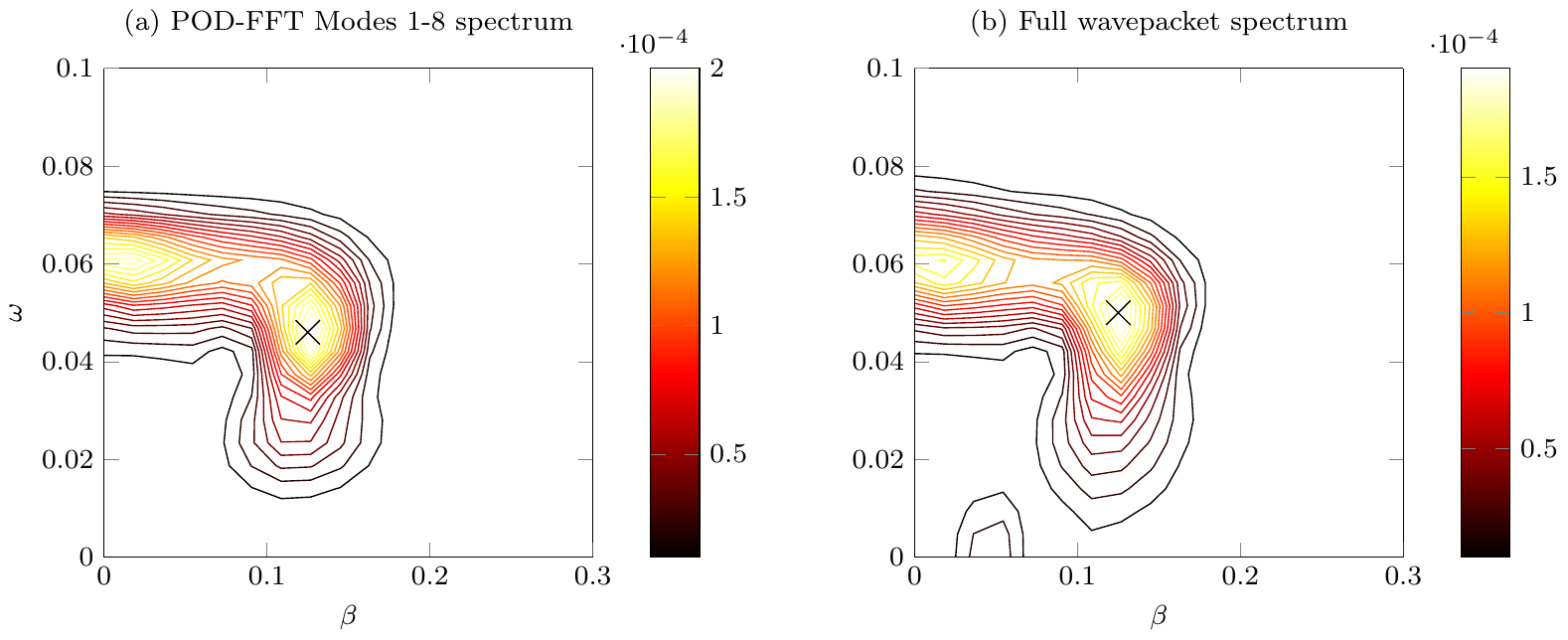}
\caption{\label{fig:single-cycle-pod-X906-fft-spacetime-combi-full} Spectral density of the combination of wavepacket POD modes 1 until 8, compared with the full wavepacket spectral density at $x=906$. The dominant oblique mode is marked ``$\times$''.}
\end{figure*}

Recall that POD modes are given by eigenfunctions $\phi_j$ that form a basis for the data set, such that the wavepacket may be expressed as a linear combination of POD modes $u_N(x)=\sum_{j=1}^{N} a_j \phi_j(x)$ (Equation~(\ref{eq:POD-finite-dimensional}) in Section~\ref{sec:PODmethod}). When all the three pairs of subharmonic POD modes (3 \& 4, 5 \& 6, 7 \& 8) are combined by the superposition of their velocities ($a_3\phi_3+a_4\phi_4+\dotsb+a_8\phi_8$), we obtain the velocity contours and spectrum in Figure~\ref{fig:single-cycle-pod-X906-fft-spacetime-combi}, which encapsulate the bulk of the subharmonic activities within the whole wavepacket. Now, the velocity contours in the central region of Figure~\ref{fig:single-cycle-pod-X906-fft-spacetime-combi}(a) look like the checkerboard pattern of a subharmonic mode, and the spectrum of the combination mode is a single oblique patch asymmetrically skewed towards the positive frequency detuning, giving it a peak frequency of $0.038$ that is greater than $\omega_0/2$, further supporting our proposal that what was previously considered as a ``single'' subharmonic mode at a poorly-understood positive frequency detuning may be better comprehended as a \emph{group} of quasi-subharmonic modes. Indeed, when the fundamental mode pair 1-2 is added to the subhamonic modes 3-8, we obtain a spectrum that is very similar to that of the full wavepacket, as in Figure~\ref{fig:single-cycle-pod-X906-fft-spacetime-combi-full}. 

\begin{figure*}    
\includegraphics{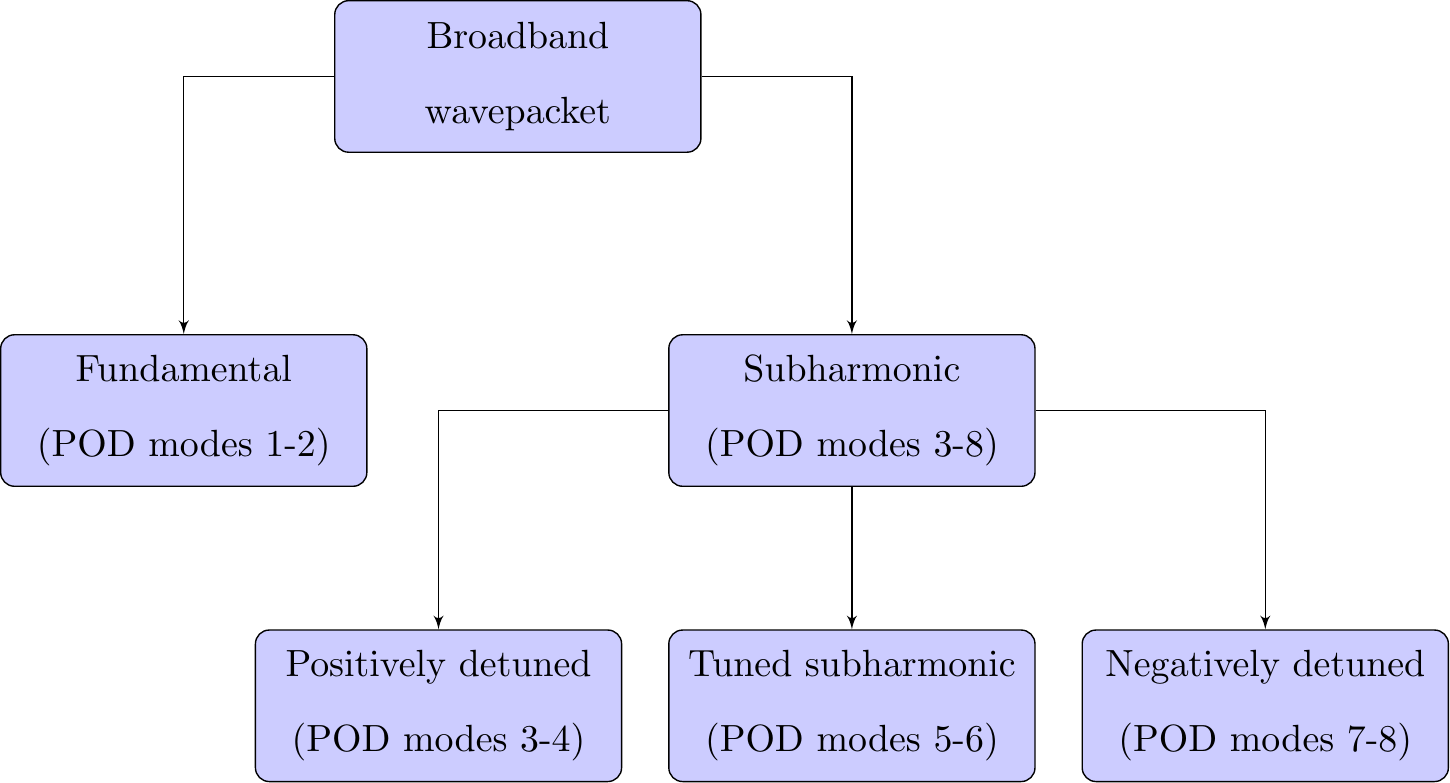}
\caption{\label{fig:POD-flowchart}Flowchart showing hierarchy of modes in a broadband wavepacket obtained by POD. The modes are numbered in decreasing order of energy, so POD Mode 1 contains the most kinetic energy.}
\end{figure*}

Our findings are summarized by the flowchart of Figure~\ref{fig:POD-flowchart}. POD modes (arranged according to their energy hierarchy) seem to reflect the progressive importance, in decreasing order, of positively detuned, tuned and negatively detuned subharmonic modes.

\subsection{\label{subsec:coherent-structures-broadband-linear-comparison} Investigating the linear/nonlinear mechanism of the POD modes}

\begin{figure*}
\includegraphics{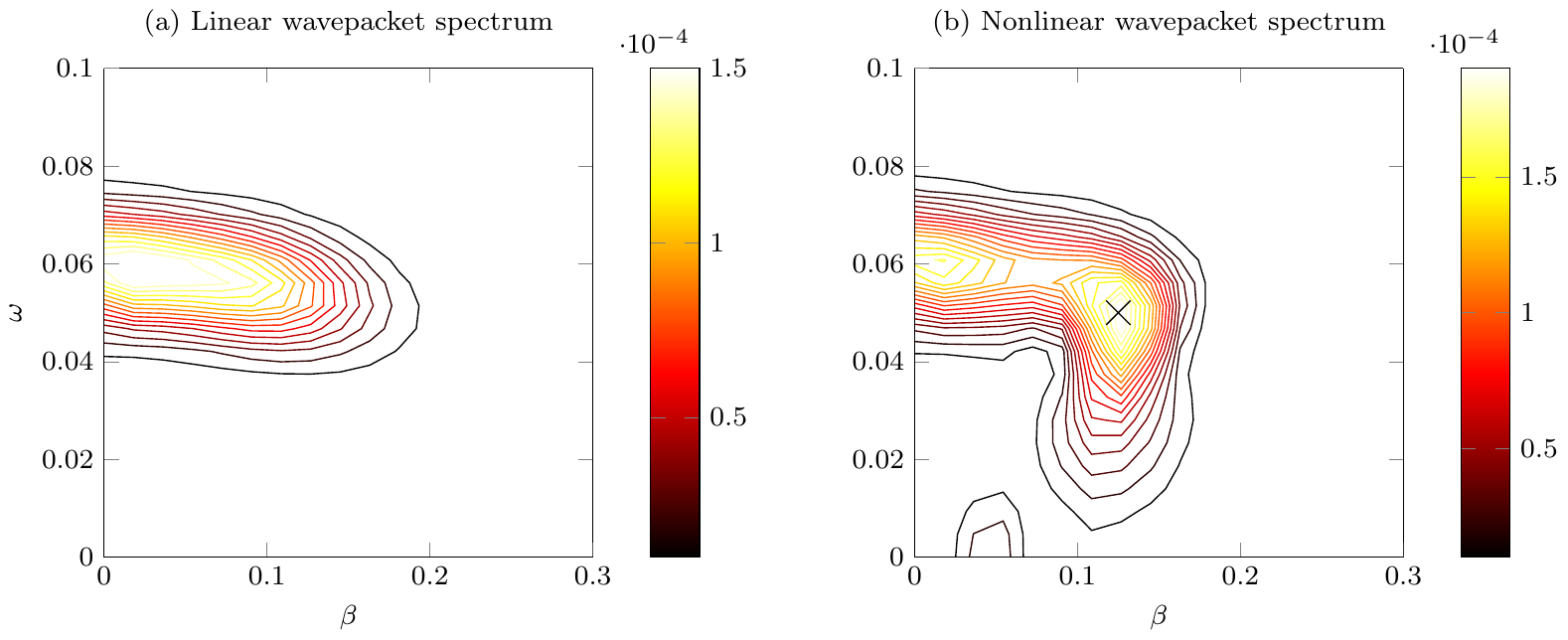}
\caption{\label{fig:single-cycle-pod-X906-fft-spacetime-linear-comparison} Spectral density of the full wavepacket at $x=906$, in the linear and nonlinear simulations. The dominant oblique mode in the nonlinear spectrum is marked ``$\times$''.}
\end{figure*}

It is of interest to determine if the POD modes are primarily a consequence of linear or nonlinear mechanisms. To investigate this matter, we first use Figure~\ref{fig:single-cycle-pod-X906-fft-spacetime-linear-comparison} to compare the spectral density contours of the wavepacket in the linear and nonlinear simulations at $x=906$. In both these simulations, the same initial source disturbance was used, but one simulation solves the nonlinear perturbation of the Navier-Stokes equations, and another solves the linear perturbation as explained in Section~\ref{subsec:dns-code}. 

Figure~\ref{fig:single-cycle-pod-X906-fft-spacetime-linear-comparison}(a) shows that the linear spectrum consists of only a single local maximum at $(\omega,\beta)\approx(0.06,0)$, corresponding to the 2D fundamental mode. On the other hand, the nonlinear spectrum in Figure~\ref{fig:single-cycle-pod-X906-fft-spacetime-linear-comparison}(b) has two local maxima, at the fundamental $(\omega,\beta)\approx(0.06,0)$ and oblique mode $(\omega,\beta)\approx(0.05,0.125)$. This already gives a good indication that the 3D POD Modes 3-8 are largely of nonlinear origin. 

\begin{figure*}
\includegraphics{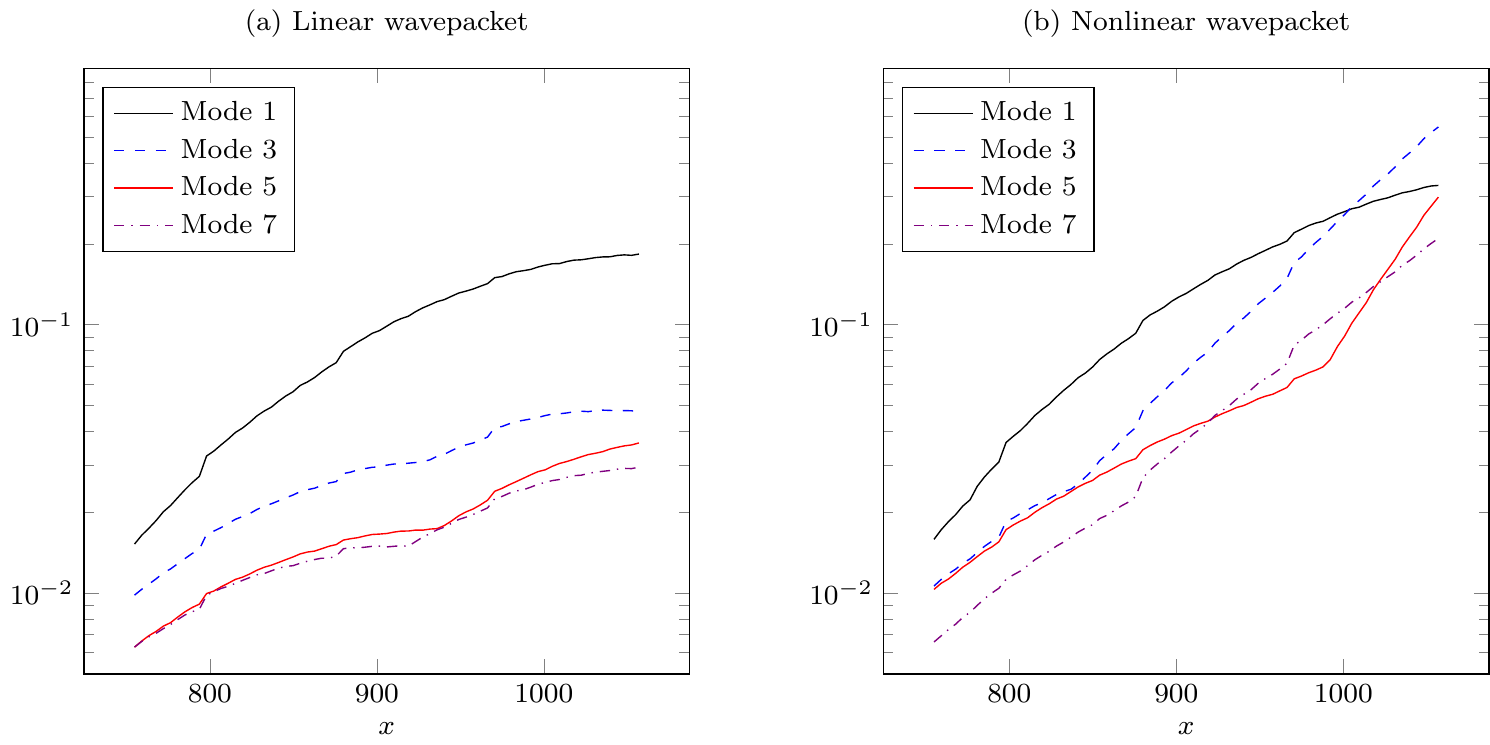}
\caption{\label{fig:single-cycle-pod-X906-expansion-coeff-projection-grouped}Normalized projection ${|(\mathbf{u}_{kn},\pmb{\phi}_j)|}^2/{\lVert\pmb{\phi}_j\rVert}^2$ of the nonlinear and linear wavepackets at various $x$-locations onto the same POD basis shown in Figure~\ref{fig:single-cycle-pod-X906-spacetime} and Figure~\ref{fig:single-cycle-pod-X906-fft-spacetime}.}
\end{figure*}

For a more quantitative comparison, we project velocity data $\mathbf{u}_{kn}$ of the linear and nonlinear wavepackets onto the POD modes 1, 3, 5 and 7 ($\pmb{\phi}_1$, $\pmb{\phi}_3$, $\pmb{\phi}_5$ and $\pmb{\phi}_7$) from the nonlinear wavepacket at $x=906$ using the inner product $(\mathbf{u}_{kn},\pmb{\phi}_j)$ (refer to Appendix~\ref{appsubsec:PODmethod-projection} for details). The projection is normalized as ${|(\mathbf{u}_{kn},\pmb{\phi}_j)|}^2/{\lVert\pmb{\phi}_j\rVert}^2$, and the result is shown in Figure~\ref{fig:single-cycle-pod-X906-expansion-coeff-projection-grouped} for the main subharmonic stage of transition. Both wavepackets are being projected onto the same nonlinear basis (which can represent both linear and nonlinear phenomena) shown in Figure~\ref{fig:single-cycle-pod-X906-spacetime} and Figure~\ref{fig:single-cycle-pod-X906-fft-spacetime}. A larger quantity ${|(\mathbf{u}_{kn},\pmb{\phi}_j)|}^2/{\lVert\pmb{\phi}_j\rVert}^2$ indicates that a given POD mode makes up a larger proportion of the overall wavepacket contents.

The fundamental POD Mode 1 is the only POD mode that shows a degree of similarity between the projections of the linear and nonlinear wavepacket in Figures \ref{fig:single-cycle-pod-X906-expansion-coeff-projection-grouped}(a) and \ref{fig:single-cycle-pod-X906-expansion-coeff-projection-grouped}(b) respectively. It may thus be deduced that POD Mode 1 has a predominantly linear character. Contrastingly, all the other Modes 3, 5 and 7 exhibit significant departure from linear growth at some point, and indicate a nonlinear mechanism at work behind them. The nonlinear growth appears to be linked to 3D spectral maxima in Figure~\ref{fig:single-cycle-pod-X906-fft-spacetime}(a) (and \ref{fig:single-cycle-pod-X906-fft}(a)) at spanwise wavenumber $\beta\approx$ 0.125, which is absent in the spectra of the linear wavepacket in Figure~\ref{fig:single-cycle-pod-X906-fft-spacetime-linear-comparison}(a). 

In terms of growth rates, the amplitude of the linear POD modes in Figure~\ref{fig:single-cycle-pod-X906-expansion-coeff-projection-grouped}(a) have a positive slope that gradually becomes flat as the wavepacket propagates downstream. This decreasing amplification factor is to be expected as the mode travels through and begins to leave the linearly unstable region of the boundary layer; whereas the nonlinear wavepacket reflects continued but weak growth beyond this point. 

In the nonlinear wavepacket of Figure~\ref{fig:single-cycle-pod-X906-expansion-coeff-projection-grouped}(b), the projection onto the fundamental Mode 1 grows similarly to the linear wavepacket, showing that this mode is only weakly affected by interactions with other 3D modes and thus plays a largely catalytic role in promoting the growth of the oblique wave modes, as noted by many researchers. The Mode 3 and 7 components of the nonlinear wavepacket can be seen to form approximate straight lines on the semi-logarithmic graph, demonstrating that they undergo steady exponential growth throughout the $x$-range shown, with the positively-detuned Mode 3 eventually overtaking the fundamental Mode 1 as the most energetic mode in the wavepacket by $x\approx1000$. On the other hand, the line for Mode 5 (Craik-type tuned resonance) has two sections, with a sharp increase in slope around $x\approx980$ signaling a shift to a higher exponential growth rate thereafter. The latter suggests that Mode 5 may eventually overtake Mode 3 as the strongest mode further downstream before wavepacket breakdown. This is supported by the findings of Yeo \textit{et al.},\cite{Yeo2010} whose extended simulation shows that the dominant wave system approaches a Craik-like tuned resonance (Mode 5), with the propagation angle of its oblique waves tending towards $60^\circ$ as imminent breakdown is approached. The latter is suggestive of an increasing inviscid (Rayleigh) character of the triad as viscous effects become secondary compared to nonlinear effects in the late stages of transition.\cite{Wu2007,Goldstein1994} 

\section{\label{conclusion} Summary and Conclusions}
In this paper, we applied the proper orthogonal decomposition (POD) technique to extract the coherent structures of a wavepacket. The largest pair of POD modes (Modes 1-2) of the wavepacket exhibit close similarities to a 2D fundamental wave, but with a shallow crescent shape and spanwise variation. The second largest pair of POD modes in the wavepacket (Modes 3-4) have an arrowhead shape often associated with a cross-section of a $\Lambda$-vortex. 

One weakness of the POD method is that the resultant POD modes are numerical or data-based eigenfunctions not easily understood within the framework of classical stability theories, such as those of Craik,\cite{Craik1971} Herbert\cite{Herbert1984,Herbert1988} and others. We are able to alleviate this difficulty by performing a FFT of POD modes to obtain their frequency-wavenumber spectrum. Using this hybrid POD-FFT, we are able to identify the dominant growing structures within the wavepacket and their corresponding spectral underpinnings. We find that energy-based POD quite remarkably extracts and distinguishes between the fundamental and dominant subharmonic modes in a wavepacket, even though it is blind to the underlying process and wavepacket physics. 

This fundamental-subharmonic resonant triad dichotomy is widely known and accepted.\cite{Craik1971,Kachanov1994} However, the POD further separates the subharmonic content of the wavepacket into three fairly distinct parts: the positively detuned mode (Mode 3), tuned Craik-type mode (Mode 5) and conjugate-detuned subharmonic modes (Mode 7), in decreasing order of energy. This distinction and hierarchy within a broadband wavepacket is less widely recognized, but it provides a possible (and much-needed) explanation for the slightly positively detuned subharmonic mode so often observed in the spectral results of previous experiments and simulations,\cite{Medeiros1999b,Yeo2010} where the subharmonic spectra have been loosely interpreted as a positively-skewed amalgamation of multiple subharmonic modes at different detuned frequencies. 

Additionally, we suggest that the positively-detuned subharmonic has the highest energy because of its preferential amplification at a rate greater than the tuned and conjugate-detuned resonances over much of the subharmonic growth stage of the wavepacket. Moreover, the broadband initial disturbance that generates the wavepacket will likely impart more energy into the positively-detuned mode by virtue of its closer spectral proximity to the fundamental frequency or mode. Future research could potentially explore energy transfer mechanisms among the modes in greater detail --- a framework for such an analysis was proposed by Dar \textit{et al.}\cite{Dar2001} and Verma.\cite{Verma2004} 

\begin{appendix}
\section{\label{appsec:POD-appendix} Details of the POD implementation}
\subsection{\label{appsubsec:PODmethod-svd} Calculating the POD via SVD}
In the finite-dimensional case, the POD can be obtained through a singular value decomposition (SVD). By defining $\mathcal{R}\phi=\langle(\phi,u)u\rangle$, we may write $\mathcal{R}\phi=\lambda\phi$. It is possible to show that  the optimal basis of a POD consists of eigenfunctions $\phi$ of the operator $\mathcal{R}$. If the flow quantities are sampled on a uniformly spaced grid and the data is represented in the form of vectors, we can use the standard inner product $(\mathbf{x},\mathbf{y})=\mathbf{y}^T\mathbf{x}$, and the linear operator $\mathcal{R}$ becomes $\mathcal{R}=\left\langle\mathbf{u}\mathbf{u}^T\right\rangle$. Hence, $\mathcal{R}\phi=\lambda\phi$ can now be expressed as
\begin{equation}
\left\langle\mathbf{u}\mathbf{u}^T\right\rangle\phi=\lambda\phi.
\label{eq:POD-eigenvalue-problem-standard-inner-product}
\end{equation}
 
The data set takes the form of a matrix $\mathbf{X}$ that can represent data in two ways, which we shall call Case A and Case B. In Case A, we have the matrix $\mathbf{X}_A$, whose columns are snapshots of the flow at successive times. Each column therefore represents the data in a streamwise-spanwise plane $u(z,x)$ at a specific time $t$ and height in the boundary layer $y$. Beginning with a 3D matrix of the velocity data, whose elements $u_{qrs}$ are a discrete sampling of the continuous function $u(t,z,x)$ on a grid with local origin at $(t_l,z_l,x_l)$, spatial spacing $\Delta_z$ and $\Delta_x$ in the $z$ and $x$ directions respectively, and time step $\Delta_t$, we convert this 3D matrix into a 2D matrix with elements $u_{kq}^A$ with the relationship
\begin{subequations}
\begin{equation}
u_{qrs}=u[t=t_l+(q-1)\Delta_t,z=z_l+(r-1)\Delta_z,x=x_l+(s-1)\Delta_x]=u_{kq}^A,
\label{eq:POD-3D-2D-matrix-conversion}
\end{equation}
\begin{equation}
k=r+(s-1)N_z.
\end{equation}
\end{subequations}
In this $u_{kq}^A$ arrangement, the first $N_z$ elements of a column correspond to the first streamwise sampling location $x=x_l$, the next $N_z$ elements in the same column correspond to the second streamwise sampling location $x=x_l+\Delta_x$, and so on. Tests indicate that the POD modes are insensitive to the arrangement of elements in the columns, as long as the mapping used to inter-convert between the column vector and matrix form is consistent. Therefore we have the $N\times M$ data matrix $\mathbf{X}_A$ with $N=(N_zN_x)$ rows and $M=N_t$ columns, 
\begin{equation}
\mathbf{X}_A=\left[\mathbf{u}_{k1}^A \dotso \mathbf{u}_{kM}^A \right].
\label{eq:data-matrix}
\end{equation}

Alternatively, we have Case B, with the data matrix $\mathbf{X}_B$, where each column represents the $u(t,z)$ data sampled at one $(x,y)$-location. In this case, we convert the 3D perturbation velocity matrix with elements $u_{qrs}$ into the 2D matrix with elements $u_{ks}^B$,
\begin{subequations}
\begin{equation}
u_{qrs}=u[t=t_l+(q-1)\Delta_t,z=z_l+(r-1)\Delta_z,x=x_l+(s-1)\Delta_x]=u_{ks}^B,
\label{eq:POD-3D-2D-matrix-conversion-tz}
\end{equation}
\begin{equation}
k=q+(r-1)N_t.
\end{equation}
\end{subequations}
In the $u_{ks}^B$ arrangement, the first $N_t$ elements in a column represent the first sampling time $t=t_l$, the next $N_t$ elements (from row $N_t+1$ until $2N_t$) are at the second sampling time $t=t_l+\Delta_t$, and so on. This leads to a $N\times M$ data matrix $\mathbf{X}_B$ where $N=(N_tN_z)$ is the number of rows and $M=N_x$ is the number of columns. Consequently, each column of $\mathbf{X}_B$ is the wavepacket data sampled at one $x$-location,
\begin{equation}
\mathbf{X}_B=\left[\mathbf{u}_{k1}^B \dotso \mathbf{u}_{kM}^B \right].
\label{eq:data-matrix-tz}
\end{equation}

Once the data matrix $\mathbf{X}_A$ or $\mathbf{X}_B$ has been formed, the subsequent steps are identical in either case, and $\mathbf{X}=\left[\mathbf{u}_{k1} \dotso \mathbf{u}_{kM} \right]$ will be used from here onwards to represent either case. Taking the arithmetic mean average of $M$ vectors, $\langle\mathbf{u}\rangle =(1/M)\sum_{n=1}^{M} \mathbf{u}_{kn}$, the eigenvalue problem (\ref{eq:POD-eigenvalue-problem-standard-inner-product}) with eigenvector $\pmb{\phi}$ is now
\begin{equation}
\frac{1}{M}\mathbf{XX}^T\pmb{\phi}=\lambda\pmb{\phi}.
\label{eq:eigenvalue-problem}
\end{equation}
Note that $\mathbf{XX}^T$ has dimension $N\times N$.

We then find the SVD of $\mathbf{X}$,
\begin{equation}
\mathbf{X}=\mathbf{Q\Sigma}\mathbf{R}^T=\sum\limits_{j=1}^{r} \sigma_j \pmb{\phi}_j \mathbf{v}_j^T 
\label{eq:eigenvalue-svd}
\end{equation}
where $\mathbf{Q}=\left[\pmb{\phi}_1 \dotso \pmb{\phi}_N \right]$ and $\mathbf{R}=\left[\mathbf{v}_1 \dotso \mathbf{v}_M \right]$ are orthogonal matrices ($\mathbf{Q}^T\mathbf{Q}=\mathbf{I}_{N\times N}$ and $\mathbf{R}^T\mathbf{R}=\mathbf{I}_{M\times M}$), $r$ is the rank of $\mathbf{X}$ and $\mathbf{\Sigma}$ is a matrix that contains the singular values $\sigma_j$ along its main diagonal, arranged in descending order. Substituting (\ref{eq:eigenvalue-svd}) into the left-hand side of (\ref{eq:eigenvalue-problem}),
\begin{equation*}
\frac{1}{M}\mathbf{XX}^T\pmb{\phi}_j
=\frac{1}{M} \mathbf{Q\Sigma}\mathbf{R}^T\mathbf{R}\mathbf{\Sigma}^T\mathbf{Q}^T\pmb{\phi}_j
=\frac{1}{M}\sigma_j^2\pmb{\phi}_j.
\end{equation*}
Comparing with (\ref{eq:eigenvalue-problem}), it can be seen that the POD modes are the columns $\pmb{\phi}_j$ of $\mathbf{Q}$ and the empirical eigenvalues are $\lambda_j=\sigma_j^2/M$. Since the singular values in the diagonal matrix $\mathbf{\Sigma}_1$ are arranged in decreasing order, Hilbert-Schmidt theory can be used to show that for velocity data, the POD modes are thus ordered in terms of kinetic energy, with the first mode $\pmb{\phi}_1$ containing the largest proportion of the total kinetic energy.

In order to express the $n$-th snapshot of the flow as a linear combination of the POD modes, $\mathbf{u}_{kn}=\sum_{j=1}^{N} a_j^n \pmb{\phi}_j$, the coefficients $a_j^n$ need to be found. This is achieved by projecting the data snapshot $\mathbf{u}_{kn}$ onto the POD mode $\pmb{\phi}_j$ using the inner product $a_j^n=(\mathbf{u}_{kn},\pmb{\phi}_j)/\lVert\pmb{\phi}_j\rVert^2$.

By calculating $\left(\sum_{j=1}^{N} a_j^n \pmb{\phi}_j\right)$ and comparing it with $\mathbf{u}_{kn}$, we may check for the convergence and accuracy of our numerical implementation of the POD. We find that the infinity norm of the difference, ${\left\lVert \mathbf{u}_{kn} -\left(\sum_{j=1}^{N} a_j^n \pmb{\phi}_j\right) \right\rVert}_\infty$ is less than $10^{-14}$ in our computational results, where the infinity norm of a vector $\mathbf{x}={(x_1,x_2,\dotsc,x_N)}^T$ is defined as $\lVert\mathbf{x}\rVert_\infty=\max\{|x_1|,|x_2|,\dotsc,|x_N|\}$.

\subsection{\label{appsubsec:PODmethod-projection} Projection of wavepackets onto a specific POD mode}
In Section~\ref{subsec:coherent-structures-broadband-linear-comparison}, projection of wavepackets onto individual modes $\pmb{\phi}_j$ is performed as ${|(\mathbf{u}_{kn},\pmb{\phi}_j)|}^2/{\lVert\pmb{\phi}_j\rVert}^2$. This produces a scalar value that represents the extent to which the POD mode $\pmb{\phi}_j$ represents that wavepacket. Among other things, it is used for understanding how a linear wavepacket is represented by a nonlinear basis. Nevertheless, because the wavepacket is moving in space and time, and the POD mode is extracted at one position in space and time, it is necessary to translate the wavepacket such that it aligns with the basis being used. Without such an alignment step, the projection ${|(\mathbf{u}_{kn},\pmb{\phi}_j)|}^2/{\lVert\pmb{\phi}_j\rVert}^2$ will be less than 1 even if the wavepacket is identical in form to $\pmb{\phi}_j$. 

The actual procedure used is as follows. Let $\mathbf{u}_{kn}^B$ be the column vector obtained when we fix $s=n$ in $u_{ks}^B$ of (\ref{eq:POD-3D-2D-matrix-conversion-tz}). In other words, $\mathbf{u}_{kn}^B$ corresponds to the $n$-th column of the matrix $\mathbf{X}_B=\left[\mathbf{u}_{k1}^B \dotso \mathbf{u}_{kM}^B \right]$ in (\ref{eq:data-matrix-tz}). The data in $\mathbf{u}_{kn}^B$ was collected at a fixed streamwise location $x=x_l+(n-1)\Delta_x$ in the time window from $t=t_l$ to $t=t_l+(N_t-1)\Delta_t$, with window center $t_c=[t_l+(N_t-1)\Delta_t]/2$. Introducing the notation $\mathbf{u}_{kn}^B(\tau)$ to refer to the wavepacket data vector $\mathbf{u}_{kn}^B$ collected in a time window of width $(N_t-1)\Delta_t$ centered at $\tau$, we form the matrix
\begin{equation}
\mathbf{F}_n=
\begin{bmatrix}
\mathbf{u}_{kn}^B(\tau=T_L)& \mathbf{u}_{kn}^B(\tau=T_L+\Delta_t)& \dotsc& \mathbf{u}_{kn}^B(\tau=T_U-\Delta_t)& \mathbf{u}_{kn}^B(\tau=T_U)
\end{bmatrix}
\end{equation}
$T_L$ is the earliest window center time, and $T_U$ is the latest window center time, with fixed $(T_L+T_U)/2=[t_l+(N_t-1)\Delta_t]/2$ and $(T_U-T_L)\geq(N_t-1)\Delta_t$. Successive columns of matrix $\mathbf{F}_n$ refer to the wavepacket viewed through windows that are displaced from each other by $\Delta_t$.

Next, we project the columns of $\mathbf{F}_n$ onto the POD modes as
\begin{equation}
\mathbf{Q}^T\mathbf{F}_n=\mathbf{G}_n
\end{equation}
where the columns of matrix $\mathbf{Q}=\left[\pmb{\phi}_1 \dotso \pmb{\phi}_N \right]$ are the POD modes $\pmb{\phi}_j$ that are the solution to an eigenproblem $\frac{1}{M}\mathbf{X}_B\mathbf{X}_B^T\pmb{\phi}=\lambda\pmb{\phi}$, as in (\ref{eq:eigenvalue-svd}). In order to find the maximum value of the wavepacket projection onto a POD mode $\pmb{\phi}_j$, we need to find the maximum element in row $j$ of the matrix $\mathbf{G}_n$. To do likewise for all the POD modes, we form the vector $\mathbf{p}_n$
\begin{equation}
\mathbf{p}_n=
\begin{pmatrix}
\max[
\begin{matrix}
\mathbf{G}_{n,1,1}& \mathbf{G}_{n,1,2}& \mathbf{G}_{n,1,3}& \dotsc& \mathbf{G}_{n,1,C}
\end{matrix}]\\

\max[
\begin{matrix}
\mathbf{G}_{n,2,1}& \mathbf{G}_{n,2,2}& \mathbf{G}_{n,2,3}& \dotsc& \mathbf{G}_{n,2,C}
\end{matrix}]\\

\vdots\\

\max[
\begin{matrix}
\mathbf{G}_{n,N,1}& \mathbf{G}_{n,N,2}& \mathbf{G}_{n,N,3}& \dotsc& \mathbf{G}_{n,N,C}
\end{matrix}]

\end{pmatrix},
\end{equation}
where $\mathbf{G}_{n,a,b}$ is the element of matrix $\mathbf{G}_n$ in row $a$ and column $b$, and $C$ is the number of columns in $\mathbf{F}_n$.

Finally, we form the matrix $\mathbf{P}$ by calculating the vector $\mathbf{p}_n$ at successive $n$, and these vectors $\mathbf{p}_n$ form the columns of $\mathbf{P}$,
\begin{equation}
\mathbf{P}=
\begin{bmatrix}
\mathbf{p}_1& \mathbf{p}_2& \mathbf{p}_3& \dotsc& \mathbf{p}_M
\end{bmatrix}.
\end{equation}
Each row $j$ of $\mathbf{P}$ is the maximum projection of the $u(t,z)$ wavepacket $\mathbf{u}_{kn}^B$ onto the POD mode $\pmb{\phi}_j$, namely $(\mathbf{u}_{kn}^B,\pmb{\phi}_j)$. The $x$-location at which the $u(t,z)$ wavepacket is sampled is determined by the column index $n$ of the matrix $\mathbf{P}$ as $x=x_l+(n-1)\Delta_x$. For instance, the line plots in Figure~\ref{fig:single-cycle-pod-X906-expansion-coeff-projection-grouped}(a) of Section~\ref{subsec:coherent-structures-broadband-linear-comparison} for the projections onto POD Mode 1, Mode 3, Mode 5 and Mode 7 correspond to the squared values of the first, third, fifth and seventh rows of matrix $\mathbf{P}$ respectively.

\end{appendix}

\bibliography{ms}

\begin{thebibliography}{65}%
\makeatletter
\providecommand \@ifxundefined [1]{%
 \@ifx{#1\undefined}
}%
\providecommand \@ifnum [1]{%
 \ifnum #1\expandafter \@firstoftwo
 \else \expandafter \@secondoftwo
 \fi
}%
\providecommand \@ifx [1]{%
 \ifx #1\expandafter \@firstoftwo
 \else \expandafter \@secondoftwo
 \fi
}%
\providecommand \natexlab [1]{#1}%
\providecommand \enquote  [1]{``#1''}%
\providecommand \bibnamefont  [1]{#1}%
\providecommand \bibfnamefont [1]{#1}%
\providecommand \citenamefont [1]{#1}%
\providecommand \href@noop [0]{\@secondoftwo}%
\providecommand \href [0]{\begingroup \@sanitize@url \@href}%
\providecommand \@href[1]{\@@startlink{#1}\@@href}%
\providecommand \@@href[1]{\endgroup#1\@@endlink}%
\providecommand \@sanitize@url [0]{\catcode `\\12\catcode `\$12\catcode
  `\&12\catcode `\#12\catcode `\^12\catcode `\_12\catcode `\%12\relax}%
\providecommand \@@startlink[1]{}%
\providecommand \@@endlink[0]{}%
\providecommand \url  [0]{\begingroup\@sanitize@url \@url }%
\providecommand \@url [1]{\endgroup\@href {#1}{\urlprefix }}%
\providecommand \urlprefix  [0]{URL }%
\providecommand \Eprint [0]{\href }%
\providecommand \doibase [0]{http://dx.doi.org/}%
\providecommand \selectlanguage [0]{\@gobble}%
\providecommand \bibinfo  [0]{\@secondoftwo}%
\providecommand \bibfield  [0]{\@secondoftwo}%
\providecommand \translation [1]{[#1]}%
\providecommand \BibitemOpen [0]{}%
\providecommand \bibitemStop [0]{}%
\providecommand \bibitemNoStop [0]{.\EOS\space}%
\providecommand \EOS [0]{\spacefactor3000\relax}%
\providecommand \BibitemShut  [1]{\csname bibitem#1\endcsname}%
\let\auto@bib@innerbib\@empty
\bibitem [{\citenamefont {Squire}(1933)}]{Squire1933}%
  \BibitemOpen
  \bibfield  {author} {\bibinfo {author} {\bibfnamefont {H.~B.}\ \bibnamefont
  {Squire}},\ }\bibfield  {title} {\enquote {\bibinfo {title} {On the stability
  for three dimensional disturbances of viscous fluid flow between parallel
  walls},}\ }\href {http://dx.doi.org/10.1098/rspa.1933.0193} {\bibfield
  {journal} {\bibinfo  {journal} {Proc. R. Soc. Lond. A}\ }\textbf {\bibinfo
  {volume} {142}},\ \bibinfo {pages} {621--628} (\bibinfo {year}
  {1933})}\BibitemShut {NoStop}%
\bibitem [{\citenamefont {Schmid}\ and\ \citenamefont
  {Henningson}(2001)}]{Schmid2001}%
  \BibitemOpen
  \bibfield  {author} {\bibinfo {author} {\bibfnamefont {P.~J.}\ \bibnamefont
  {Schmid}}\ and\ \bibinfo {author} {\bibfnamefont {D.~S.}\ \bibnamefont
  {Henningson}},\ }\href@noop {} {\emph {\bibinfo {title} {Stability and
  Transition in Shear Flows}}},\ Vol.\ \bibinfo {volume} {142}\ (\bibinfo
  {publisher} {Springer},\ \bibinfo {year} {2001})\BibitemShut {NoStop}%
\bibitem [{\citenamefont {Morkovin}(1969)}]{Morkovin1969}%
  \BibitemOpen
  \bibfield  {author} {\bibinfo {author} {\bibfnamefont {M.~V.}\ \bibnamefont
  {Morkovin}},\ }\enquote {\bibinfo {title} {On the many faces of
  transition},}\ in\ \href {\doibase 10.1007/978-1-4899-5579-1_1} {\emph
  {\bibinfo {booktitle} {Viscous Drag Reduction}}},\ \bibinfo {editor} {edited
  by\ \bibinfo {editor} {\bibfnamefont {C.~S.}\ \bibnamefont {Wells}}}\
  (\bibinfo  {publisher} {Springer US},\ \bibinfo {address} {Boston, MA},\
  \bibinfo {year} {1969})\ pp.\ \bibinfo {pages} {1--31}\BibitemShut {NoStop}%
\bibitem [{\citenamefont {Schubauer}\ and\ \citenamefont
  {Skramstad}(1947)}]{Schubauer1947}%
  \BibitemOpen
  \bibfield  {author} {\bibinfo {author} {\bibfnamefont {G.~B.}\ \bibnamefont
  {Schubauer}}\ and\ \bibinfo {author} {\bibfnamefont {H.~K.}\ \bibnamefont
  {Skramstad}},\ }\bibfield  {title} {\enquote {\bibinfo {title} {Laminar
  boundary-layer oscillations and transition on a flat plate},}\ }\href
  {http://dx.doi.org/10.6028/jres.038.013} {\bibfield  {journal} {\bibinfo
  {journal} {J. Res. Natl. Bur. Stand.}\ }\textbf {\bibinfo {volume} {38}},\
  \bibinfo {pages} {251--292} (\bibinfo {year} {1947})}\BibitemShut {NoStop}%
\bibitem [{\citenamefont {Klebanoff}, \citenamefont {Tidstrom},\ and\
  \citenamefont {Sargent}(1962)}]{Klebanoff1962}%
  \BibitemOpen
  \bibfield  {author} {\bibinfo {author} {\bibfnamefont {P.~S.}\ \bibnamefont
  {Klebanoff}}, \bibinfo {author} {\bibfnamefont {K.~D.}\ \bibnamefont
  {Tidstrom}}, \ and\ \bibinfo {author} {\bibfnamefont {L.~M.}\ \bibnamefont
  {Sargent}},\ }\bibfield  {title} {\enquote {\bibinfo {title} {The
  three-dimensional nature of boundary-layer instability},}\ }\href
  {http://dx.doi.org/10.1017/S0022112062000014} {\bibfield  {journal} {\bibinfo
   {journal} {J. Fluid Mech.}\ }\textbf {\bibinfo {volume} {12}},\ \bibinfo
  {pages} {1--34} (\bibinfo {year} {1962})}\BibitemShut {NoStop}%
\bibitem [{\citenamefont {Fasel}(2002)}]{Fasel2002}%
  \BibitemOpen
  \bibfield  {author} {\bibinfo {author} {\bibfnamefont {H.~F.}\ \bibnamefont
  {Fasel}},\ }\bibfield  {title} {\enquote {\bibinfo {title} {Numerical
  investigation of the interaction of the {Klebanoff-mode} with a
  {Tollmien-Schlichting} wave},}\ }\href
  {http://dx.doi.org/10.1017/S0022112002006140} {\bibfield  {journal} {\bibinfo
   {journal} {J. Fluid Mech.}\ }\textbf {\bibinfo {volume} {450}},\ \bibinfo
  {pages} {1--33} (\bibinfo {year} {2002})}\BibitemShut {NoStop}%
\bibitem [{\citenamefont {Sengupta}\ and\ \citenamefont
  {Bhaumik}(2011)}]{Sengupta2011}%
  \BibitemOpen
  \bibfield  {author} {\bibinfo {author} {\bibfnamefont {T.~K.}\ \bibnamefont
  {Sengupta}}\ and\ \bibinfo {author} {\bibfnamefont {S.}~\bibnamefont
  {Bhaumik}},\ }\bibfield  {title} {\enquote {\bibinfo {title} {Onset of
  turbulence from the receptivity stage of fluid flows},}\ }\href
  {http://dx.doi.org/10.1103/PhysRevLett.107.154501} {\bibfield  {journal}
  {\bibinfo  {journal} {Phys. Rev. Lett.}\ }\textbf {\bibinfo {volume} {107}},\
  \bibinfo {pages} {154501} (\bibinfo {year} {2011})}\BibitemShut {NoStop}%
\bibitem [{\citenamefont {Sengupta}, \citenamefont {Bhaumik},\ and\
  \citenamefont {Bhumkar}(2012)}]{Sengupta2012}%
  \BibitemOpen
  \bibfield  {author} {\bibinfo {author} {\bibfnamefont {T.~K.}\ \bibnamefont
  {Sengupta}}, \bibinfo {author} {\bibfnamefont {S.}~\bibnamefont {Bhaumik}}, \
  and\ \bibinfo {author} {\bibfnamefont {Y.~G.}\ \bibnamefont {Bhumkar}},\
  }\bibfield  {title} {\enquote {\bibinfo {title} {Direct numerical simulation
  of two-dimensional wall-bounded turbulent flows from receptivity stage},}\
  }\href {\doibase http://dx.doi.org/10.1103/PhysRevE.85.026308} {\bibfield
  {journal} {\bibinfo  {journal} {Phys. Rev. E}\ }\textbf {\bibinfo {volume}
  {85}},\ \bibinfo {pages} {026308} (\bibinfo {year} {2012})}\BibitemShut
  {NoStop}%
\bibitem [{\citenamefont {Bhaumik}\ and\ \citenamefont
  {Sengupta}(2014)}]{Bhaumik2014a}%
  \BibitemOpen
  \bibfield  {author} {\bibinfo {author} {\bibfnamefont {S.}~\bibnamefont
  {Bhaumik}}\ and\ \bibinfo {author} {\bibfnamefont {T.~K.}\ \bibnamefont
  {Sengupta}},\ }\bibfield  {title} {\enquote {\bibinfo {title} {Precursor of
  transition to turbulence: Spatiotemporal wave front},}\ }\href
  {http://dx.doi.org/10.1103/PhysRevE.89.043018} {\bibfield  {journal}
  {\bibinfo  {journal} {Phys. Rev. E}\ }\textbf {\bibinfo {volume} {89}},\
  \bibinfo {pages} {043018} (\bibinfo {year} {2014})}\BibitemShut {NoStop}%
\bibitem [{\citenamefont {Saric}, \citenamefont {Reed},\ and\ \citenamefont
  {Kerschen}(2002)}]{Saric2002}%
  \BibitemOpen
  \bibfield  {author} {\bibinfo {author} {\bibfnamefont {W.~S.}\ \bibnamefont
  {Saric}}, \bibinfo {author} {\bibfnamefont {H.~L.}\ \bibnamefont {Reed}}, \
  and\ \bibinfo {author} {\bibfnamefont {E.~J.}\ \bibnamefont {Kerschen}},\
  }\bibfield  {title} {\enquote {\bibinfo {title} {Boundary-layer receptivity
  to freestream disturbances},}\ }\href {\doibase
  10.1146/annurev.fluid.34.082701.161921} {\bibfield  {journal} {\bibinfo
  {journal} {Annu. Rev. Fluid Mech.}\ }\textbf {\bibinfo {volume} {34}},\
  \bibinfo {pages} {291--319} (\bibinfo {year} {2002})}\BibitemShut {NoStop}%
\bibitem [{\citenamefont {Grosch}\ and\ \citenamefont
  {Salwen}(1978)}]{Grosch1978}%
  \BibitemOpen
  \bibfield  {author} {\bibinfo {author} {\bibfnamefont {C.~E.}\ \bibnamefont
  {Grosch}}\ and\ \bibinfo {author} {\bibfnamefont {H.}~\bibnamefont
  {Salwen}},\ }\bibfield  {title} {\enquote {\bibinfo {title} {The continuous
  spectrum of the orr-sommerfeld equation. part 1. the spectrum and the
  eigenfunctions},}\ }\href {http://dx.doi.org/10.1017/S0022112078002918}
  {\bibfield  {journal} {\bibinfo  {journal} {J. Fluid Mech.}\ }\textbf
  {\bibinfo {volume} {87}},\ \bibinfo {pages} {33--54} (\bibinfo {year}
  {1978})}\BibitemShut {NoStop}%
\bibitem [{\citenamefont {Liu}, \citenamefont {Zaki},\ and\ \citenamefont
  {Durbin}(2008)}]{Liu2008}%
  \BibitemOpen
  \bibfield  {author} {\bibinfo {author} {\bibfnamefont {Y.}~\bibnamefont
  {Liu}}, \bibinfo {author} {\bibfnamefont {T.~A.}\ \bibnamefont {Zaki}}, \
  and\ \bibinfo {author} {\bibfnamefont {P.~A.}\ \bibnamefont {Durbin}},\
  }\bibfield  {title} {\enquote {\bibinfo {title} {Boundary-layer transition by
  interaction of discrete and continuous modes},}\ }\href
  {http://dx.doi.org/10.1017/S0022112008001201} {\bibfield  {journal} {\bibinfo
   {journal} {J. Fluid Mech.}\ }\textbf {\bibinfo {volume} {604}},\ \bibinfo
  {pages} {199--233} (\bibinfo {year} {2008})}\BibitemShut {NoStop}%
\bibitem [{\citenamefont {Craik}(1971)}]{Craik1971}%
  \BibitemOpen
  \bibfield  {author} {\bibinfo {author} {\bibfnamefont {A.~D.~D.}\
  \bibnamefont {Craik}},\ }\bibfield  {title} {\enquote {\bibinfo {title}
  {Non-linear resonant instability in boundary layers},}\ }\href
  {http://dx.doi.org/10.1017/S0022112071002635} {\bibfield  {journal} {\bibinfo
   {journal} {J. Fluid Mech.}\ }\textbf {\bibinfo {volume} {50}},\ \bibinfo
  {pages} {393--413} (\bibinfo {year} {1971})}\BibitemShut {NoStop}%
\bibitem [{\citenamefont {Herbert}(1984)}]{Herbert1984}%
  \BibitemOpen
  \bibfield  {author} {\bibinfo {author} {\bibfnamefont {T.}~\bibnamefont
  {Herbert}},\ }\bibfield  {title} {\enquote {\bibinfo {title} {Analysis of the
  subharmonic route to transition in boundary layers},}\ }\href
  {http://dx.doi.org/10.2514/6.1984-9} {\bibfield  {journal} {\bibinfo
  {journal} {AIAA Paper 1984-0009}\ } (\bibinfo {year} {1984})}\BibitemShut
  {NoStop}%
\bibitem [{\citenamefont {Herbert}(1988)}]{Herbert1988}%
  \BibitemOpen
  \bibfield  {author} {\bibinfo {author} {\bibfnamefont {T.}~\bibnamefont
  {Herbert}},\ }\bibfield  {title} {\enquote {\bibinfo {title} {Secondary
  instability of boundary layers},}\ }\href {\doibase
  10.1146/annurev.fl.20.010188.002415} {\bibfield  {journal} {\bibinfo
  {journal} {Annu. Rev. Fluid Mech.}\ }\textbf {\bibinfo {volume} {20}},\
  \bibinfo {pages} {487--526} (\bibinfo {year} {1988})}\BibitemShut {NoStop}%
\bibitem [{\citenamefont {Kachanov}\ and\ \citenamefont
  {Levchenko}(1984)}]{Kachanov1984}%
  \BibitemOpen
  \bibfield  {author} {\bibinfo {author} {\bibfnamefont {Y.~S.}\ \bibnamefont
  {Kachanov}}\ and\ \bibinfo {author} {\bibfnamefont {V.~Y.}\ \bibnamefont
  {Levchenko}},\ }\bibfield  {title} {\enquote {\bibinfo {title} {The resonant
  interaction of disturbances at laminar-turbulent transition in a boundary
  layer},}\ }\href {http://dx.doi.org/10.1017/S0022112084000100} {\bibfield
  {journal} {\bibinfo  {journal} {J. Fluid Mech.}\ }\textbf {\bibinfo {volume}
  {138}},\ \bibinfo {pages} {209--247} (\bibinfo {year} {1984})}\BibitemShut
  {NoStop}%
\bibitem [{\citenamefont {Saric}(1984)}]{Saric1984}%
  \BibitemOpen
  \bibfield  {author} {\bibinfo {author} {\bibfnamefont {W.}~\bibnamefont
  {Saric}},\ }\bibfield  {title} {\enquote {\bibinfo {title} {Forced and
  unforced subharmonic resonance in boundary-layer transition},}\ }\href
  {http://dx.doi.org/10.2514/6.1984-7} {\bibfield  {journal} {\bibinfo
  {journal} {AIAA Paper 1984-0007}\ } (\bibinfo {year} {1984})}\BibitemShut
  {NoStop}%
\bibitem [{\citenamefont {Boiko}\ \emph {et~al.}(2012)\citenamefont {Boiko},
  \citenamefont {Dovgal}, \citenamefont {Grek},\ and\ \citenamefont
  {Kozlov}}]{Boiko2012}%
  \BibitemOpen
  \bibfield  {author} {\bibinfo {author} {\bibfnamefont {A.~V.}\ \bibnamefont
  {Boiko}}, \bibinfo {author} {\bibfnamefont {A.~V.}\ \bibnamefont {Dovgal}},
  \bibinfo {author} {\bibfnamefont {G.~R.}\ \bibnamefont {Grek}}, \ and\
  \bibinfo {author} {\bibfnamefont {V.~V.}\ \bibnamefont {Kozlov}},\ }\enquote
  {\bibinfo {title} {Physics of transitional shear flows},}\ \ (\bibinfo
  {publisher} {Springer},\ \bibinfo {year} {2012})\ Chap.~\bibinfo {chapter}
  {12}, pp.\ \bibinfo {pages} {225--226}\BibitemShut {NoStop}%
\bibitem [{\citenamefont {Kachanov}(1994)}]{Kachanov1994}%
  \BibitemOpen
  \bibfield  {author} {\bibinfo {author} {\bibfnamefont {Y.~S.}\ \bibnamefont
  {Kachanov}},\ }\bibfield  {title} {\enquote {\bibinfo {title} {Physical
  mechanisms of laminar-boundary-layer transition},}\ }\href {\doibase
  doi:10.1146/annurev.fl.26.010194.002211} {\bibfield  {journal} {\bibinfo
  {journal} {Annu. Rev. Fluid Mech.}\ }\textbf {\bibinfo {volume} {26}},\
  \bibinfo {pages} {411--482} (\bibinfo {year} {1994})}\BibitemShut {NoStop}%
\bibitem [{\citenamefont {Borodulin}\ \emph {et~al.}(2002)\citenamefont
  {Borodulin}, \citenamefont {Kachanov}, \citenamefont {Koptsev},\ and\
  \citenamefont {Roschektayev}}]{Borodulin2002b}%
  \BibitemOpen
  \bibfield  {author} {\bibinfo {author} {\bibfnamefont {V.~I.}\ \bibnamefont
  {Borodulin}}, \bibinfo {author} {\bibfnamefont {Y.~S.}\ \bibnamefont
  {Kachanov}}, \bibinfo {author} {\bibfnamefont {D.~B.}\ \bibnamefont
  {Koptsev}}, \ and\ \bibinfo {author} {\bibfnamefont {A.~P.}\ \bibnamefont
  {Roschektayev}},\ }\bibfield  {title} {\enquote {\bibinfo {title}
  {Experimental study of resonant interactions of instability waves in
  self-similar boundary layer with an adverse pressure gradient: {II}.
  {Detuned} resonances},}\ }\href {http://dx.doi.org/10.1088/1468-5248/3/1/063}
  {\bibfield  {journal} {\bibinfo  {journal} {J. Turbul.}\ }\textbf {\bibinfo
  {volume} {3}},\ \bibinfo {pages} {N63} (\bibinfo {year} {2002})}\BibitemShut
  {NoStop}%
\bibitem [{\citenamefont {W{\"{u}}rz}\ \emph {et~al.}(2012)\citenamefont
  {W{\"{u}}rz}, \citenamefont {Sartorius}, \citenamefont {Kloker},
  \citenamefont {Borodulin}, \citenamefont {Kachanov},\ and\ \citenamefont
  {Smorodsky}}]{Wurz2012a}%
  \BibitemOpen
  \bibfield  {author} {\bibinfo {author} {\bibfnamefont {W.}~\bibnamefont
  {W{\"{u}}rz}}, \bibinfo {author} {\bibfnamefont {D.}~\bibnamefont
  {Sartorius}}, \bibinfo {author} {\bibfnamefont {M.}~\bibnamefont {Kloker}},
  \bibinfo {author} {\bibfnamefont {V.~I.}\ \bibnamefont {Borodulin}}, \bibinfo
  {author} {\bibfnamefont {Y.~S.}\ \bibnamefont {Kachanov}}, \ and\ \bibinfo
  {author} {\bibfnamefont {B.~V.}\ \bibnamefont {Smorodsky}},\ }\bibfield
  {title} {\enquote {\bibinfo {title} {Detuned resonances of
  {Tollmien--Schlichting} waves in an airfoil boundary layer: Experiment,
  theory, and direct numerical simulation},}\ }\href {\doibase
  http://dx.doi.org/10.1063/1.4751246} {\bibfield  {journal} {\bibinfo
  {journal} {Phys. Fluids}\ }\textbf {\bibinfo {volume} {24}},\ \bibinfo
  {pages} {094103--25} (\bibinfo {year} {2012})}\BibitemShut {NoStop}%
\bibitem [{\citenamefont {Wu}, \citenamefont {Stewart},\ and\ \citenamefont
  {Cowley}(2007)}]{Wu2007}%
  \BibitemOpen
  \bibfield  {author} {\bibinfo {author} {\bibfnamefont {X.}~\bibnamefont
  {Wu}}, \bibinfo {author} {\bibfnamefont {P.~A.}\ \bibnamefont {Stewart}}, \
  and\ \bibinfo {author} {\bibfnamefont {S.~J.}\ \bibnamefont {Cowley}},\
  }\bibfield  {title} {\enquote {\bibinfo {title} {On the catalytic role of the
  phase-locked interaction of {Tollmien-Schlichting} waves in boundary-layer
  transition},}\ }\href {\doibase 10.1017/S002211200700804X} {\bibfield
  {journal} {\bibinfo  {journal} {J. Fluid Mech.}\ }\textbf {\bibinfo {volume}
  {590}},\ \bibinfo {pages} {265--294} (\bibinfo {year} {2007})}\BibitemShut
  {NoStop}%
\bibitem [{\citenamefont {Borodulin}, \citenamefont {Kachanov},\ and\
  \citenamefont {Koptsev}(2002)}]{Borodulin2002c}%
  \BibitemOpen
  \bibfield  {author} {\bibinfo {author} {\bibfnamefont {V.~I.}\ \bibnamefont
  {Borodulin}}, \bibinfo {author} {\bibfnamefont {Y.~S.}\ \bibnamefont
  {Kachanov}}, \ and\ \bibinfo {author} {\bibfnamefont {D.~B.}\ \bibnamefont
  {Koptsev}},\ }\bibfield  {title} {\enquote {\bibinfo {title} {Experimental
  study of resonant interactions of instability waves in self-similar boundary
  layer with an adverse pressure gradient: {III}. {Broadband} disturbances},}\
  }\href {http://dx.doi.org/10.1088/1468-5248/3/1/064} {\bibfield  {journal}
  {\bibinfo  {journal} {J. Turbul.}\ }\textbf {\bibinfo {volume} {3}},\
  \bibinfo {pages} {N64} (\bibinfo {year} {2002})}\BibitemShut {NoStop}%
\bibitem [{\citenamefont {Sengupta}, \citenamefont {Rao},\ and\ \citenamefont
  {Venkatasubbaiah}(2006{\natexlab{a}})}]{Sengupta2006}%
  \BibitemOpen
  \bibfield  {author} {\bibinfo {author} {\bibfnamefont {T.~K.}\ \bibnamefont
  {Sengupta}}, \bibinfo {author} {\bibfnamefont {A.~K.}\ \bibnamefont {Rao}}, \
  and\ \bibinfo {author} {\bibfnamefont {K.}~\bibnamefont {Venkatasubbaiah}},\
  }\bibfield  {title} {\enquote {\bibinfo {title} {Spatiotemporal growing wave
  fronts in spatially stable boundary layers},}\ }\href
  {http://dx.doi.org/10.1103/PhysRevLett.96.224504} {\bibfield  {journal}
  {\bibinfo  {journal} {Phys. Rev. Lett.}\ }\textbf {\bibinfo {volume} {96}},\
  \bibinfo {pages} {224504} (\bibinfo {year} {2006}{\natexlab{a}})}\BibitemShut
  {NoStop}%
\bibitem [{\citenamefont {Sengupta}, \citenamefont {Rao},\ and\ \citenamefont
  {Venkatasubbaiah}(2006{\natexlab{b}})}]{Sengupta2006a}%
  \BibitemOpen
  \bibfield  {author} {\bibinfo {author} {\bibfnamefont {T.~K.}\ \bibnamefont
  {Sengupta}}, \bibinfo {author} {\bibfnamefont {A.~K.}\ \bibnamefont {Rao}}, \
  and\ \bibinfo {author} {\bibfnamefont {K.}~\bibnamefont {Venkatasubbaiah}},\
  }\bibfield  {title} {\enquote {\bibinfo {title} {Spatio-temporal growth of
  disturbances in a boundary layer and energy based receptivity analysis},}\
  }\href {\doibase http://dx.doi.org/10.1063/1.2348732} {\bibfield  {journal}
  {\bibinfo  {journal} {Phys. Fluids}\ }\textbf {\bibinfo {volume} {18}},\
  \bibinfo {pages} {094101} (\bibinfo {year} {2006}{\natexlab{b}})}\BibitemShut
  {NoStop}%
\bibitem [{\citenamefont {Breuer}, \citenamefont {Cohen},\ and\ \citenamefont
  {Haritonidis}(1997)}]{Breuer1997}%
  \BibitemOpen
  \bibfield  {author} {\bibinfo {author} {\bibfnamefont {K.~S.}\ \bibnamefont
  {Breuer}}, \bibinfo {author} {\bibfnamefont {J.}~\bibnamefont {Cohen}}, \
  and\ \bibinfo {author} {\bibfnamefont {J.~H.}\ \bibnamefont {Haritonidis}},\
  }\bibfield  {title} {\enquote {\bibinfo {title} {The late stages of
  transition induced by a low-amplitude wavepacket in a laminar boundary
  layer},}\ }\href {http://dx.doi.org/10.1017/S0022112097005417} {\bibfield
  {journal} {\bibinfo  {journal} {J. Fluid Mech.}\ }\textbf {\bibinfo {volume}
  {340}},\ \bibinfo {pages} {395--411} (\bibinfo {year} {1997})}\BibitemShut
  {NoStop}%
\bibitem [{\citenamefont {Medeiros}\ and\ \citenamefont
  {Gaster}(1999{\natexlab{a}})}]{Medeiros1999a}%
  \BibitemOpen
  \bibfield  {author} {\bibinfo {author} {\bibfnamefont {M.~A.~F.}\
  \bibnamefont {Medeiros}}\ and\ \bibinfo {author} {\bibfnamefont
  {M.}~\bibnamefont {Gaster}},\ }\bibfield  {title} {\enquote {\bibinfo {title}
  {The influence of phase on the nonlinear evolution of wavepackets in boundary
  layers},}\ }\href {http://dx.doi.org/10.1017/S0022112099006175} {\bibfield
  {journal} {\bibinfo  {journal} {J. Fluid Mech.}\ }\textbf {\bibinfo {volume}
  {397}},\ \bibinfo {pages} {259--283} (\bibinfo {year}
  {1999}{\natexlab{a}})}\BibitemShut {NoStop}%
\bibitem [{\citenamefont {Medeiros}\ and\ \citenamefont
  {Gaster}(1999{\natexlab{b}})}]{Medeiros1999b}%
  \BibitemOpen
  \bibfield  {author} {\bibinfo {author} {\bibfnamefont {M.~A.~F.}\
  \bibnamefont {Medeiros}}\ and\ \bibinfo {author} {\bibfnamefont
  {M.}~\bibnamefont {Gaster}},\ }\bibfield  {title} {\enquote {\bibinfo {title}
  {The production of subharmonic waves in the nonlinear evolution of
  wavepackets in boundary layers},}\ }\href
  {http://dx.doi.org/10.1017/S0022112099006424} {\bibfield  {journal} {\bibinfo
   {journal} {J. Fluid Mech.}\ }\textbf {\bibinfo {volume} {399}},\ \bibinfo
  {pages} {301--318} (\bibinfo {year} {1999}{\natexlab{b}})}\BibitemShut
  {NoStop}%
\bibitem [{\citenamefont {Yeo}\ \emph {et~al.}(2010)\citenamefont {Yeo},
  \citenamefont {Zhao}, \citenamefont {Wang},\ and\ \citenamefont
  {Ng}}]{Yeo2010}%
  \BibitemOpen
  \bibfield  {author} {\bibinfo {author} {\bibfnamefont {K.~S.}\ \bibnamefont
  {Yeo}}, \bibinfo {author} {\bibfnamefont {X.}~\bibnamefont {Zhao}}, \bibinfo
  {author} {\bibfnamefont {Z.~Y.}\ \bibnamefont {Wang}}, \ and\ \bibinfo
  {author} {\bibfnamefont {K.~C.}\ \bibnamefont {Ng}},\ }\bibfield  {title}
  {\enquote {\bibinfo {title} {{DNS} of wavepacket evolution in a {Blasius}
  boundary layer},}\ }\href {http://dx.doi.org/10.1017/S0022112009994095}
  {\bibfield  {journal} {\bibinfo  {journal} {J. Fluid Mech.}\ }\textbf
  {\bibinfo {volume} {652}},\ \bibinfo {pages} {333--372} (\bibinfo {year}
  {2010})}\BibitemShut {NoStop}%
\bibitem [{\citenamefont {Cohen}, \citenamefont {Breuer},\ and\ \citenamefont
  {Haritonidis}(1991)}]{Cohen1991}%
  \BibitemOpen
  \bibfield  {author} {\bibinfo {author} {\bibfnamefont {J.}~\bibnamefont
  {Cohen}}, \bibinfo {author} {\bibfnamefont {K.~S.}\ \bibnamefont {Breuer}}, \
  and\ \bibinfo {author} {\bibfnamefont {J.~H.}\ \bibnamefont {Haritonidis}},\
  }\bibfield  {title} {\enquote {\bibinfo {title} {On the evolution of a wave
  packet in a laminar boundary layer},}\ }\href
  {http://dx.doi.org/10.1017/S0022112091002185} {\bibfield  {journal} {\bibinfo
   {journal} {J. Fluid Mech.}\ }\textbf {\bibinfo {volume} {225}},\ \bibinfo
  {pages} {575--606} (\bibinfo {year} {1991})}\BibitemShut {NoStop}%
\bibitem [{\citenamefont {Wang}(2003)}]{Wang2003}%
  \BibitemOpen
  \bibfield  {author} {\bibinfo {author} {\bibfnamefont {Z.~Y.}\ \bibnamefont
  {Wang}},\ }\emph {\bibinfo {title} {Computational simulation of unsteady
  boundary layer over compliant surfaces}},\ \href
  {http://scholarbank.nus.sg/handle/10635/13874} {Ph.D. thesis},\ \bibinfo
  {school} {National University of Singapore} (\bibinfo {year}
  {2003})\BibitemShut {NoStop}%
\bibitem [{\citenamefont {Wang}, \citenamefont {Yeo},\ and\ \citenamefont
  {Khoo}(2005)}]{Wang2005}%
  \BibitemOpen
  \bibfield  {author} {\bibinfo {author} {\bibfnamefont {Z.~Y.}\ \bibnamefont
  {Wang}}, \bibinfo {author} {\bibfnamefont {K.~S.}\ \bibnamefont {Yeo}}, \
  and\ \bibinfo {author} {\bibfnamefont {B.~C.}\ \bibnamefont {Khoo}},\
  }\bibfield  {title} {\enquote {\bibinfo {title} {Spatial direct numerical
  simulation of transitional boundary layer over compliant surfaces},}\ }\href
  {\doibase http://dx.doi.org/10.1016/j.compfluid.2004.08.005} {\bibfield
  {journal} {\bibinfo  {journal} {Comput. Fluids}\ }\textbf {\bibinfo {volume}
  {34}},\ \bibinfo {pages} {1062--1095} (\bibinfo {year} {2005})}\BibitemShut
  {NoStop}%
\bibitem [{\citenamefont {Sengupta}\ \emph {et~al.}(2015)\citenamefont
  {Sengupta}, \citenamefont {Sathyanarayanan}, \citenamefont {Sriramkrishnan},\
  and\ \citenamefont {Mulloth}}]{Sengupta2015b}%
  \BibitemOpen
  \bibfield  {author} {\bibinfo {author} {\bibfnamefont {T.~K.}\ \bibnamefont
  {Sengupta}}, \bibinfo {author} {\bibfnamefont {V.~K.}\ \bibnamefont
  {Sathyanarayanan}}, \bibinfo {author} {\bibfnamefont {M.}~\bibnamefont
  {Sriramkrishnan}}, \ and\ \bibinfo {author} {\bibfnamefont {A.}~\bibnamefont
  {Mulloth}},\ }\bibfield  {title} {\enquote {\bibinfo {title} {Role of time
  integration in computing transitional flows caused by wall excitation},}\
  }\href {\doibase http://dx.doi.org/10.1007/s10915-014-9967-1} {\bibfield
  {journal} {\bibinfo  {journal} {Journal of Scientific Computing}\ }\textbf
  {\bibinfo {volume} {65}},\ \bibinfo {pages} {224--248} (\bibinfo {year}
  {2015})}\BibitemShut {NoStop}%
\bibitem [{\citenamefont {Sengupta}(2016)}]{Sengupta2016}%
  \BibitemOpen
  \bibfield  {author} {\bibinfo {author} {\bibfnamefont {T.~K.}\ \bibnamefont
  {Sengupta}},\ }\enquote {\bibinfo {title} {A critical assessment of
  simulations for transitional and turbulent flows},}\ in\ \href@noop {} {\emph
  {\bibinfo {booktitle} {Advances in Computation, Modeling and Control of
  Transitional and Turbulent Flows}}},\ \bibinfo {editor} {edited by\ \bibinfo
  {editor} {\bibfnamefont {T.~K.}\ \bibnamefont {Sengupta}}, \bibinfo {editor}
  {\bibfnamefont {S.~K.}\ \bibnamefont {Lele}}, \bibinfo {editor}
  {\bibfnamefont {K.~R.}\ \bibnamefont {Sreenivasan}}, \ and\ \bibinfo {editor}
  {\bibfnamefont {P.~A.}\ \bibnamefont {Davidson}}}\ (\bibinfo  {publisher}
  {World Scientific},\ \bibinfo {address} {Singapore},\ \bibinfo {year}
  {2016})\BibitemShut {NoStop}%
\bibitem [{\citenamefont {Chorin}(1969)}]{Chorin1969}%
  \BibitemOpen
  \bibfield  {author} {\bibinfo {author} {\bibfnamefont {A.~J.}\ \bibnamefont
  {Chorin}},\ }\bibfield  {title} {\enquote {\bibinfo {title} {On the
  convergence of discrete approximations to the {Navier-Stokes} equations},}\
  }\href {http://dx.doi.org/10.1090/S0025-5718-1969-0242393-5} {\bibfield
  {journal} {\bibinfo  {journal} {Math. Comp.}\ }\textbf {\bibinfo {volume}
  {23}},\ \bibinfo {pages} {341--353} (\bibinfo {year} {1969})}\BibitemShut
  {NoStop}%
\bibitem [{\citenamefont {Temam}(1984)}]{Temam1984}%
  \BibitemOpen
  \bibfield  {author} {\bibinfo {author} {\bibfnamefont {R.}~\bibnamefont
  {Temam}},\ }\href@noop {} {\emph {\bibinfo {title} {{Navier-Stokes}
  Equations: Theory and Numerical Analysis}}},\ Vol.\ \bibinfo {volume} {343}\
  (\bibinfo  {publisher} {AMS Chelsea Publishing},\ \bibinfo {year} {1984})\
  p.\ \bibinfo {pages} {408}\BibitemShut {NoStop}%
\bibitem [{\citenamefont {Kim}\ and\ \citenamefont {Moin}(1985)}]{Kim1985}%
  \BibitemOpen
  \bibfield  {author} {\bibinfo {author} {\bibfnamefont {J.}~\bibnamefont
  {Kim}}\ and\ \bibinfo {author} {\bibfnamefont {P.}~\bibnamefont {Moin}},\
  }\bibfield  {title} {\enquote {\bibinfo {title} {Application of a
  fractional-step method to incompressible {Navier-Stokes} equations},}\ }\href
  {\doibase http://dx.doi.org/10.1016/0021-9991(85)90148-2} {\bibfield
  {journal} {\bibinfo  {journal} {J. Comput. Phys.}\ }\textbf {\bibinfo
  {volume} {59}},\ \bibinfo {pages} {308--323} (\bibinfo {year}
  {1985})}\BibitemShut {NoStop}%
\bibitem [{\citenamefont {Rhie}\ and\ \citenamefont {Chow}(1983)}]{Rhie1983}%
  \BibitemOpen
  \bibfield  {author} {\bibinfo {author} {\bibfnamefont {C.~M.}\ \bibnamefont
  {Rhie}}\ and\ \bibinfo {author} {\bibfnamefont {W.~L.}\ \bibnamefont
  {Chow}},\ }\bibfield  {title} {\enquote {\bibinfo {title} {Numerical study of
  the turbulent flow past an airfoil with trailing edge separation},}\ }\href
  {http://dx.doi.org/10.2514/3.8284} {\bibfield  {journal} {\bibinfo  {journal}
  {AIAA J.}\ }\textbf {\bibinfo {volume} {21}},\ \bibinfo {pages} {1525--1532}
  (\bibinfo {year} {1983})}\BibitemShut {NoStop}%
\bibitem [{\citenamefont {Wesseling}\ and\ \citenamefont
  {Oosterlee}(2001)}]{Wesseling2001}%
  \BibitemOpen
  \bibfield  {author} {\bibinfo {author} {\bibfnamefont {P.}~\bibnamefont
  {Wesseling}}\ and\ \bibinfo {author} {\bibfnamefont {C.~W.}\ \bibnamefont
  {Oosterlee}},\ }\bibfield  {title} {\enquote {\bibinfo {title} {Geometric
  multigrid with applications to computational fluid dynamics},}\ }\href
  {\doibase http://dx.doi.org/10.1016/S0377-0427(00)00517-3} {\bibfield
  {journal} {\bibinfo  {journal} {J. Comput. Appl. Math.}\ }\textbf {\bibinfo
  {volume} {128}},\ \bibinfo {pages} {311--334} (\bibinfo {year}
  {2001})}\BibitemShut {NoStop}%
\bibitem [{\citenamefont {Birkhoff}, \citenamefont {Varga},\ and\ \citenamefont
  {Young}(1962)}]{Birkhoff1962}%
  \BibitemOpen
  \bibfield  {author} {\bibinfo {author} {\bibfnamefont {G.}~\bibnamefont
  {Birkhoff}}, \bibinfo {author} {\bibfnamefont {R.~S.}\ \bibnamefont {Varga}},
  \ and\ \bibinfo {author} {\bibfnamefont {D.}~\bibnamefont {Young}},\
  }\enquote {\bibinfo {title} {Alternating direction implicit methods},}\ in\
  \href {\doibase http://dx.doi.org/10.1016/S0065-2458(08)60620-8} {\emph
  {\bibinfo {booktitle} {Advances in Computers}}},\ Vol.~\bibinfo {volume}
  {3},\ \bibinfo {editor} {edited by\ \bibinfo {editor} {\bibfnamefont {L.~A.}\
  \bibnamefont {Franz}}\ and\ \bibinfo {editor} {\bibfnamefont
  {R.}~\bibnamefont {Morris}}}\ (\bibinfo  {publisher} {Elsevier},\ \bibinfo
  {year} {1962})\ pp.\ \bibinfo {pages} {189--273}\BibitemShut {NoStop}%
\bibitem [{\citenamefont {Brandt}(1977)}]{Brandt1977}%
  \BibitemOpen
  \bibfield  {author} {\bibinfo {author} {\bibfnamefont {A.}~\bibnamefont
  {Brandt}},\ }\bibfield  {title} {\enquote {\bibinfo {title} {Multi-level
  adaptive solutions to boundary-value problems},}\ }\href
  {http://dx.doi.org/10.1090/S0025-5718-1977-0431719-X} {\bibfield  {journal}
  {\bibinfo  {journal} {Math. Comp.}\ }\textbf {\bibinfo {volume} {31}},\
  \bibinfo {pages} {333--390} (\bibinfo {year} {1977})}\BibitemShut {NoStop}%
\bibitem [{\citenamefont {Liu}\ and\ \citenamefont {Liu}(1994)}]{Liu1994}%
  \BibitemOpen
  \bibfield  {author} {\bibinfo {author} {\bibfnamefont {Z.}~\bibnamefont
  {Liu}}\ and\ \bibinfo {author} {\bibfnamefont {C.}~\bibnamefont {Liu}},\
  }\bibfield  {title} {\enquote {\bibinfo {title} {Fourth order finite
  difference and multigrid methods for modeling instabilities in flat plate
  boundary layers\textemdash{2-D} and {3-D} approaches},}\ }\href {\doibase
  http://dx.doi.org/10.1016/0045-7930(94)90063-9} {\bibfield  {journal}
  {\bibinfo  {journal} {Comput. Fluids}\ }\textbf {\bibinfo {volume} {23}},\
  \bibinfo {pages} {955--982} (\bibinfo {year} {1994})}\BibitemShut {NoStop}%
\bibitem [{\citenamefont {Kang}\ and\ \citenamefont {Yeo}(2013)}]{Kang2013}%
  \BibitemOpen
  \bibfield  {author} {\bibinfo {author} {\bibfnamefont {K.~L.}\ \bibnamefont
  {Kang}}\ and\ \bibinfo {author} {\bibfnamefont {K.~S.}\ \bibnamefont {Yeo}},\
  }\bibfield  {title} {\enquote {\bibinfo {title} {The effect of wavepacket
  frequency bandwidth on the laminar-turbulent transition process in a
  {Blasius} boundary layer},}\ }\href {\doibase
  http://dx.doi.org/10.2514/6.2013-2615} {\bibfield  {journal} {\bibinfo
  {journal} {AIAA Paper 2013-2615}\ } (\bibinfo {year} {2013}),\
  http://dx.doi.org/10.2514/6.2013-2615}\BibitemShut {NoStop}%
\bibitem [{\citenamefont {Kang}\ and\ \citenamefont {Yeo}(2015)}]{Kang2015}%
  \BibitemOpen
  \bibfield  {author} {\bibinfo {author} {\bibfnamefont {K.-L.}\ \bibnamefont
  {Kang}}\ and\ \bibinfo {author} {\bibfnamefont {K.~S.}\ \bibnamefont {Yeo}},\
  }\bibfield  {title} {\enquote {\bibinfo {title} {The combined effects of
  wavepacket frequency, amplitude and bandwidth on its transition process in a
  boundary layer},}\ }\href {\doibase
  http://dx.doi.org/10.1016/j.piutam.2015.03.061} {\bibfield  {journal}
  {\bibinfo  {journal} {Procedia IUTAM}\ }\textbf {\bibinfo {volume} {14}},\
  \bibinfo {pages} {364--373} (\bibinfo {year} {2015})}\BibitemShut {NoStop}%
\bibitem [{\citenamefont {Craik}(2001)}]{Craik2001}%
  \BibitemOpen
  \bibfield  {author} {\bibinfo {author} {\bibfnamefont {A.~D.~D.}\
  \bibnamefont {Craik}},\ }\bibfield  {title} {\enquote {\bibinfo {title} {A
  model for subharmonic resonance within wavepackets in unstable boundary
  layers},}\ }\href {http://dx.doi.org/10.1017/S0022112001003524} {\bibfield
  {journal} {\bibinfo  {journal} {J. Fluid Mech.}\ }\textbf {\bibinfo {volume}
  {432}},\ \bibinfo {pages} {409--418} (\bibinfo {year} {2001})}\BibitemShut
  {NoStop}%
\bibitem [{\citenamefont {Zhao}(2007)}]{Zhao2007}%
  \BibitemOpen
  \bibfield  {author} {\bibinfo {author} {\bibfnamefont {X.}~\bibnamefont
  {Zhao}},\ }\emph {\bibinfo {title} {Computational simulation of wavepacket
  evolution over compliant surfaces}},\ \href
  {http://scholarbank.nus.edu.sg/handle/10635/16155} {Ph.D. thesis},\ \bibinfo
  {school} {National University of Singapore} (\bibinfo {year}
  {2007})\BibitemShut {NoStop}%
\bibitem [{\citenamefont {Fasel}, \citenamefont {Rist},\ and\ \citenamefont
  {Konzelmann}(1990)}]{Fasel1990b}%
  \BibitemOpen
  \bibfield  {author} {\bibinfo {author} {\bibfnamefont {H.~F.}\ \bibnamefont
  {Fasel}}, \bibinfo {author} {\bibfnamefont {U.}~\bibnamefont {Rist}}, \ and\
  \bibinfo {author} {\bibfnamefont {U.}~\bibnamefont {Konzelmann}},\ }\bibfield
   {title} {\enquote {\bibinfo {title} {Numerical investigation of the
  three-dimensional development in boundary-layer transition},}\ }\href
  {\doibase http://dx.doi.org/10.2514/3.10349} {\bibfield  {journal} {\bibinfo
  {journal} {AIAA J.}\ }\textbf {\bibinfo {volume} {28}},\ \bibinfo {pages}
  {29--37} (\bibinfo {year} {1990})}\BibitemShut {NoStop}%
\bibitem [{\citenamefont {Liu}\ and\ \citenamefont {Liu}(1995)}]{Liu1995}%
  \BibitemOpen
  \bibfield  {author} {\bibinfo {author} {\bibfnamefont {C.}~\bibnamefont
  {Liu}}\ and\ \bibinfo {author} {\bibfnamefont {Z.}~\bibnamefont {Liu}},\
  }\bibfield  {title} {\enquote {\bibinfo {title} {Multigrid mapping and box
  relaxation for simulation of the whole process of flow transition in 3d
  boundary layers},}\ }\href {\doibase
  http://dx.doi.org/10.1006/jcph.1995.1138} {\bibfield  {journal} {\bibinfo
  {journal} {J. Comput. Phys.}\ }\textbf {\bibinfo {volume} {119}},\ \bibinfo
  {pages} {325--341} (\bibinfo {year} {1995})}\BibitemShut {NoStop}%
\bibitem [{\citenamefont {Paul}\ and\ \citenamefont {Verma}(2016)}]{Paul2016}%
  \BibitemOpen
  \bibfield  {author} {\bibinfo {author} {\bibfnamefont {S.}~\bibnamefont
  {Paul}}\ and\ \bibinfo {author} {\bibfnamefont {M.~K.}\ \bibnamefont
  {Verma}},\ }\enquote {\bibinfo {title} {Proper orthogonal decomposition vs.
  fourier analysis for extraction of large-scale structures of thermal
  convection},}\ in\ \href {https://arxiv.org/abs/1705.09897} {\emph {\bibinfo
  {booktitle} {Advances in Computation, Modeling and Control of Transitional
  and Turbulent Flows}}}\ (\bibinfo  {publisher} {World Scientific},\ \bibinfo
  {address} {Singapore},\ \bibinfo {year} {2016})\BibitemShut {NoStop}%
\bibitem [{\citenamefont {Sengupta}, \citenamefont {Swagata},\ and\
  \citenamefont {Yogesh}(2011)}]{Sengupta2011a}%
  \BibitemOpen
  \bibfield  {author} {\bibinfo {author} {\bibfnamefont {T.~K.}\ \bibnamefont
  {Sengupta}}, \bibinfo {author} {\bibfnamefont {B.}~\bibnamefont {Swagata}}, \
  and\ \bibinfo {author} {\bibfnamefont {B.}~\bibnamefont {Yogesh}},\
  }\bibfield  {title} {\enquote {\bibinfo {title} {Nonlinear receptivity and
  instability studies by proper orthogonal decomposition},}\ }\href
  {http://dx.doi.org/10.2514/6.2011-3293} {\bibfield  {journal} {\bibinfo
  {journal} {AIAA Paper 2011-3293}\ } (\bibinfo {year} {2011})}\BibitemShut
  {NoStop}%
\bibitem [{\citenamefont {Sengupta}, \citenamefont {Singh},\ and\ \citenamefont
  {Suman}(2010)}]{Sengupta2010}%
  \BibitemOpen
  \bibfield  {author} {\bibinfo {author} {\bibfnamefont {T.~K.}\ \bibnamefont
  {Sengupta}}, \bibinfo {author} {\bibfnamefont {N.}~\bibnamefont {Singh}}, \
  and\ \bibinfo {author} {\bibfnamefont {V.~K.}\ \bibnamefont {Suman}},\
  }\bibfield  {title} {\enquote {\bibinfo {title} {Dynamical system approach to
  instability of flow past a circular cylinder},}\ }\href
  {http://dx.doi.org/10.1017/S0022112010001035} {\bibfield  {journal} {\bibinfo
   {journal} {J. Fluid Mech.}\ }\textbf {\bibinfo {volume} {656}},\ \bibinfo
  {pages} {82--115} (\bibinfo {year} {2010})}\BibitemShut {NoStop}%
\bibitem [{\citenamefont {Sengupta}, \citenamefont {Vijay},\ and\ \citenamefont
  {Singh}(2011)}]{Sengupta2011b}%
  \BibitemOpen
  \bibfield  {author} {\bibinfo {author} {\bibfnamefont {T.~K.}\ \bibnamefont
  {Sengupta}}, \bibinfo {author} {\bibfnamefont {V.~V. S.~N.}\ \bibnamefont
  {Vijay}}, \ and\ \bibinfo {author} {\bibfnamefont {N.}~\bibnamefont
  {Singh}},\ }\bibfield  {title} {\enquote {\bibinfo {title} {Universal
  instability modes in internal and external flows},}\ }\href {\doibase
  http://dx.doi.org/10.1016/j.compfluid.2010.09.006} {\bibfield  {journal}
  {\bibinfo  {journal} {Comput. Fluids}\ }\textbf {\bibinfo {volume} {40}},\
  \bibinfo {pages} {221--235} (\bibinfo {year} {2011})}\BibitemShut {NoStop}%
\bibitem [{\citenamefont {Weisstein}(2015)}]{Weisstein2015}%
  \BibitemOpen
  \bibfield  {author} {\bibinfo {author} {\bibfnamefont {E.~W.}\ \bibnamefont
  {Weisstein}},\ }\href {http://mathworld.wolfram.com/FourierTransform.html}
  {\enquote {\bibinfo {title} {Fourier {Transform.} {From} {MathWorld} -- a
  {Wolfram} web resource},}\ } (\bibinfo {year} {2015})\BibitemShut {NoStop}%
\bibitem [{\citenamefont {Newland}(1993)}]{Newland1993}%
  \BibitemOpen
  \bibfield  {author} {\bibinfo {author} {\bibfnamefont {D.~E.}\ \bibnamefont
  {Newland}},\ }\href@noop {} {\emph {\bibinfo {title} {An Introduction to
  Random Vibrations, Spectral and Wavelet Analysis}}},\ \bibinfo {edition}
  {3rd}\ ed.\ (\bibinfo  {publisher} {Dover Publications, Inc},\ \bibinfo
  {address} {Mineola, New York},\ \bibinfo {year} {1993})\BibitemShut {NoStop}%
\bibitem [{\citenamefont {Kay}\ and\ \citenamefont {Marple}(1981)}]{Kay1981}%
  \BibitemOpen
  \bibfield  {author} {\bibinfo {author} {\bibfnamefont {S.~M.}\ \bibnamefont
  {Kay}}\ and\ \bibinfo {author} {\bibfnamefont {J.}~\bibnamefont {Marple},
  \bibfnamefont {S.~L.}},\ }\bibfield  {title} {\enquote {\bibinfo {title}
  {Spectrum analysis-a modern perspective},}\ }\href
  {http://dx.doi.org/10.1109/PROC.1981.12184} {\bibfield  {journal} {\bibinfo
  {journal} {Proc. IEEE}\ }\textbf {\bibinfo {volume} {69}},\ \bibinfo {pages}
  {1380--1419} (\bibinfo {year} {1981})}\BibitemShut {NoStop}%
\bibitem [{\citenamefont {Holmes}\ \emph {et~al.}(2012)\citenamefont {Holmes},
  \citenamefont {Lumley}, \citenamefont {Berkhooz},\ and\ \citenamefont
  {Rowley}}]{Holmes2012}%
  \BibitemOpen
  \bibfield  {author} {\bibinfo {author} {\bibfnamefont {P.}~\bibnamefont
  {Holmes}}, \bibinfo {author} {\bibfnamefont {J.}~\bibnamefont {Lumley}},
  \bibinfo {author} {\bibfnamefont {G.}~\bibnamefont {Berkhooz}}, \ and\
  \bibinfo {author} {\bibfnamefont {C.~W.}\ \bibnamefont {Rowley}},\
  }\href@noop {} {\emph {\bibinfo {title} {Turbulence, Coherent Structures,
  Dynamical Systems and Symmetry}}}\ (\bibinfo  {publisher} {Cambridge
  University Press},\ \bibinfo {year} {2012})\BibitemShut {NoStop}%
\bibitem [{\citenamefont {Harris}(1978)}]{Harris1978}%
  \BibitemOpen
  \bibfield  {author} {\bibinfo {author} {\bibfnamefont {F.~J.}\ \bibnamefont
  {Harris}},\ }\bibfield  {title} {\enquote {\bibinfo {title} {On the use of
  windows for harmonic analysis with the discrete {Fourier} transform},}\
  }\href {http://dx.doi.org/10.1109/PROC.1978.10837} {\bibfield  {journal}
  {\bibinfo  {journal} {Proc. IEEE}\ }\textbf {\bibinfo {volume} {66}},\
  \bibinfo {pages} {51--83} (\bibinfo {year} {1978})}\BibitemShut {NoStop}%
\bibitem [{\citenamefont {Rempfer}\ and\ \citenamefont
  {Fasel}(1994)}]{Rempfer1994b}%
  \BibitemOpen
  \bibfield  {author} {\bibinfo {author} {\bibfnamefont {D.}~\bibnamefont
  {Rempfer}}\ and\ \bibinfo {author} {\bibfnamefont {H.~F.}\ \bibnamefont
  {Fasel}},\ }\bibfield  {title} {\enquote {\bibinfo {title} {Evolution of
  three-dimensional coherent structures in a flat-plate boundary layer},}\
  }\href {http://dx.doi.org/10.1017/S0022112094003551} {\bibfield  {journal}
  {\bibinfo  {journal} {J. Fluid Mech.}\ }\textbf {\bibinfo {volume} {260}},\
  \bibinfo {pages} {351--375} (\bibinfo {year} {1994})}\BibitemShut {NoStop}%
\bibitem [{\citenamefont {Rempfer}(1994)}]{Rempfer1994a}%
  \BibitemOpen
  \bibfield  {author} {\bibinfo {author} {\bibfnamefont {D.}~\bibnamefont
  {Rempfer}},\ }\bibfield  {title} {\enquote {\bibinfo {title} {On the
  structure of dynamical systems describing the evolution of coherent
  structures in a convective boundary layer},}\ }\href {\doibase
  http://dx.doi.org/10.1063/1.868253} {\bibfield  {journal} {\bibinfo
  {journal} {Phys. Fluids}\ }\textbf {\bibinfo {volume} {6}},\ \bibinfo {pages}
  {1402--1404} (\bibinfo {year} {1994})}\BibitemShut {NoStop}%
\bibitem [{\citenamefont {Pierce}, \citenamefont {Moin},\ and\ \citenamefont
  {Sayadi}(2013)}]{Pierce2013}%
  \BibitemOpen
  \bibfield  {author} {\bibinfo {author} {\bibfnamefont {B.}~\bibnamefont
  {Pierce}}, \bibinfo {author} {\bibfnamefont {P.}~\bibnamefont {Moin}}, \ and\
  \bibinfo {author} {\bibfnamefont {T.}~\bibnamefont {Sayadi}},\ }\bibfield
  {title} {\enquote {\bibinfo {title} {Application of vortex identification
  schemes to direct numerical simulation data of a transitional boundary
  layer},}\ }\href {\doibase http://dx.doi.org/10.1063/1.4774340} {\bibfield
  {journal} {\bibinfo  {journal} {Phys. Fluids}\ }\textbf {\bibinfo {volume}
  {25}},\ \bibinfo {pages} {015102--14} (\bibinfo {year} {2013})}\BibitemShut
  {NoStop}%
\bibitem [{\citenamefont {Liang}\ and\ \citenamefont {Dong}(2015)}]{Liang2015}%
  \BibitemOpen
  \bibfield  {author} {\bibinfo {author} {\bibfnamefont {Z.}~\bibnamefont
  {Liang}}\ and\ \bibinfo {author} {\bibfnamefont {H.}~\bibnamefont {Dong}},\
  }\bibfield  {title} {\enquote {\bibinfo {title} {On the symmetry of proper
  orthogonal decomposition modes of a low-aspect-ratio plate},}\ }\href
  {\doibase http://dx.doi.org/10.1063/1.4921843} {\bibfield  {journal}
  {\bibinfo  {journal} {Phys. Fluids}\ }\textbf {\bibinfo {volume} {27}},\
  \bibinfo {pages} {063601} (\bibinfo {year} {2015})}\BibitemShut {NoStop}%
\bibitem [{\citenamefont {Ichihashi}, \citenamefont {Jeng},\ and\ \citenamefont
  {Cohen}(2010)}]{Ichihashi2010}%
  \BibitemOpen
  \bibfield  {author} {\bibinfo {author} {\bibfnamefont {F.}~\bibnamefont
  {Ichihashi}}, \bibinfo {author} {\bibfnamefont {S.-M.}\ \bibnamefont {Jeng}},
  \ and\ \bibinfo {author} {\bibfnamefont {K.}~\bibnamefont {Cohen}},\
  }\bibfield  {title} {\enquote {\bibinfo {title} {Proper orthogonal
  decomposition and fourier analysis on the energy release rate dynamics},}\
  }\href {http://dx.doi.org/10.2514/6.2010-22} {\bibfield  {journal} {\bibinfo
  {journal} {AIAA Paper 2010-22}\ } (\bibinfo {year} {2010})}\BibitemShut
  {NoStop}%
\bibitem [{\citenamefont {Goldstein}(1994)}]{Goldstein1994}%
  \BibitemOpen
  \bibfield  {author} {\bibinfo {author} {\bibfnamefont {M.~E.}\ \bibnamefont
  {Goldstein}},\ }\bibfield  {title} {\enquote {\bibinfo {title} {Nonlinear
  interactions between oblique instability waves on nearly parallel shear
  flows},}\ }\href {\doibase 10.1063/1.868311} {\bibfield  {journal} {\bibinfo
  {journal} {Phys. Fluids}\ }\textbf {\bibinfo {volume} {6}},\ \bibinfo {pages}
  {724--735} (\bibinfo {year} {1994})}\BibitemShut {NoStop}%
\bibitem [{\citenamefont {Dar}, \citenamefont {Verma},\ and\ \citenamefont
  {Eswaran}(2001)}]{Dar2001}%
  \BibitemOpen
  \bibfield  {author} {\bibinfo {author} {\bibfnamefont {G.}~\bibnamefont
  {Dar}}, \bibinfo {author} {\bibfnamefont {M.~K.}\ \bibnamefont {Verma}}, \
  and\ \bibinfo {author} {\bibfnamefont {V.}~\bibnamefont {Eswaran}},\
  }\bibfield  {title} {\enquote {\bibinfo {title} {Energy transfer in
  two-dimensional magnetohydrodynamic turbulence: formalism and numerical
  results},}\ }\href {\doibase http://dx.doi.org/10.1016/S0167-2789(01)00307-4}
  {\bibfield  {journal} {\bibinfo  {journal} {Phys. D}\ }\textbf {\bibinfo
  {volume} {157}},\ \bibinfo {pages} {207--225} (\bibinfo {year}
  {2001})}\BibitemShut {NoStop}%
\bibitem [{\citenamefont {Verma}(2004)}]{Verma2004}%
  \BibitemOpen
  \bibfield  {author} {\bibinfo {author} {\bibfnamefont {M.~K.}\ \bibnamefont
  {Verma}},\ }\bibfield  {title} {\enquote {\bibinfo {title} {Statistical
  theory of magnetohydrodynamic turbulence: recent results},}\ }\href {\doibase
  http://dx.doi.org/10.1016/j.physrep.2004.07.007} {\bibfield  {journal}
  {\bibinfo  {journal} {Phys. Rep.}\ }\textbf {\bibinfo {volume} {401}},\
  \bibinfo {pages} {229--380} (\bibinfo {year} {2004})}\BibitemShut {NoStop}%
\end{thebibliography}%


%

\end{document}